\documentclass[fleqn,usenatbib]{mnras}
\hypersetup{linkcolor=blue,citecolor=blue,filecolor=cyan,urlcolor=magenta}

\usepackage{newtxtext,newtxmath}
\usepackage[T1]{fontenc}
\usepackage{graphicx}
\usepackage{amsmath}
\usepackage{bm}

\title[Phase-Resolved RVM fitting to AR Sco Data]{Probing the Non-thermal Emission Geometry of AR Sco via Optical 
Phase-Resolved Polarimetry}

\author[L. Du Plessis et al.]{
Louis du Plessis,$^{1}$\thanks{E-mail: louisdp95@gmail.com}
Christo Venter,$^{1}$
Zorawar Wadiasingh,$^{1,2,3,4}$
Alice K. Harding,$^{5}$
David A. H. Buckley,$^{6}$
Stephen B. Potter$^{6}$
and P. J. Meintjes$^{7}$
\\
$^{1}$Centre for Space Research, North-West University, Private Bag X6001, Potchefstroom 2520, South Africa\\
$^{2}$Astrophysics Science Division, NASA Goddard Space Flight Center, Greenbelt, MD 20771, USA\\
$^{3}$Department of Astronomy, University of Maryland, College Park, Maryland 20742, USA\\
$^{4}$Center for Research and Exploration in Space Science and Technology, NASA/GSFC, Greenbelt, Maryland 20771, USA\\
$^{5}$Theoretical Division, Los Alamos National Laboratory, Los Alamos, NM 58545, USA\\
$^{6}$South African Astronomical Observatory, PO Box 9, Observatory, 7935, Cape Town, South Africa\\
$^{7}$Department of Physics, University of the Free State, PO Box 339, Bloemfontein, South Africa, 930
}

\date{Accepted 2021 December 3. Received 2021 December 3; in original form 2021 September 23}
\pubyear{2021}

\begin{document}
\label{firstpage}
\pagerange{\pageref{firstpage}--\pageref{lastpage}}
\maketitle

\begin{abstract} 
AR Sco is a binary system that contains a white and red dwarf. The rotation rate of the white dwarf has been observed to slow down, analogous to rotation-powered radio pulsars; it has thus been dubbed a ``white dwarf pulsar''. We previously fit the traditional radio pulsar rotating vector model to the linearly polarised optical data from this source, constraining the system geometry as well as the white dwarf mass. Using a much more extensive dataset, we now explore the application of the same model to binary phase-resolved optical polarimetric data, thought to be the result of non-thermal synchrotron radiation, and derive the magnetic inclination angle $\alpha$ and the observer angle $\zeta$ at different orbital phases. We obtain a $\sim 10^{\circ}$ variation in $\alpha$ and $\sim 30^{\circ}$ variation in $\zeta$ over the orbital period. The variation patterns in these two parameters is robust, regardless of the binning and epoch of data used. We speculate that the observer is detecting radiation from an asymmetric emission region that is a stable structure over several orbital periods. The success of this simple model lastly implies that the pitch angles of the particles are small and the pulsed, non-thermal emission originates relatively close to the white dwarf surface.        
\end{abstract} 


\section{Introduction} \label{intro}
AR Scorpii (AR Sco) was recently reclassified as a novel binary system  containing a white dwarf (WD) and an M-dwarf companion \citep{Marsh2016}, based on the detection of pulsed non-thermal emission extending from radio to X-rays \citep{Buckley2017,Takata2018}. The majority of proposed models (Section~\ref{sec:4}) propose that this emission is due to synchrotron radiation (SR). Analysis of \textit{Fermi} LAT data by \cite{Kaplan2019} indicated a possible detection at $3.8\sigma$ significance, which may point to the existence of an additional spectral component. More recently, however, and using more than 10 years of \textit{Fermi} Large Area Telescope (LAT) data, \cite{Singh2020} derived a 2$\sigma$ upper limit of $2.27\times 10^{-12}$~erg~cm$^{-2}$~s$^{-1}$ on the integral energy flux above 100 MeV from the binary system AR Sco. 
 
A binary orbital period of $P_{\rm b} =3.56$ hours, a WD spin period of $P= 1.95$ minutes, and a beat period of 1.97 minutes were inferred  \citep{Marsh2016}. Interestingly, the orbital separation $a$ is smaller than the WD light cylinder radius\footnote{This is the radius where the rotation speed equals that of light in vacuum.} $R_{\rm LC}$, implying that the companion is located inside the WD magnetosphere at $a \approx 0.13 R_{\rm LC}$. The novelty of the system is due to an established WD spin-down rate first inferred by \citet{Marsh2016} and more firmly established by  \citet{Stiller2018} using  extensive observations as $\dot{P} = 7.18\times10^{-13}\,\rm{\, s \, s^{-1}}$. The implied  spin-down power $\dot{E}_{\rm rot} = I_{\rm WD} \Omega \dot{\Omega} \approx 5 \times 10^{33}$ erg s$^{-1}$, with $\Omega$ the angular velocity and $\dot{\Omega}$ its time derivative, for a fiducial WD moment of inertia $ I_{\rm WD} = 3 \times 10^{50}$ g cm${}^2$, indicates that the observed non-thermal emission can indeed be powered by the rotational spin-down of the WD. 

The uniqueness of this system was confirmed by the observed lack of Doppler-broadened emission lines from accreting gas, suggesting the absence of an accretion disk\footnote{Conversely, \citet{Takata2018} possibly detected an iron ionisation emission line using \textit{XMM} Newton. This was confirmed by \textit{NuStar} observations \citep{Koglin2005}. Such a line could be indicative of the presence of an accretion disk, but this does not fit the observed broadband spectrum. Also, one could speculate that the iron line is due to ionised material in the system.}. This is supported by the fact that the X-ray luminosity is only 4\% of the total optically-dominated luminosity and is only $\sim1\%$ of the X-ray luminosities of typical intermediate polars (IP), and also from the fact that all optical and ultraviolet emission lines originate from the irradiated face of the M-dwarf companion.
This means that AR Sco has different properties from standard IPs and is not in a propeller state as the well-known AR Aquarii. 

The dual characteristics of observed WD spin-down and absence of an accretion disk led \cite{Buckley2017} to attribute the observed non-thermal luminosity to magnetic dipole radiation\footnote{\label{footnote_3} The idea of magnetic dipole radiation dates back to the original picture of a pulsar magnetic field rotating in a vacuum. The frequency of this radiation is coupled to the rotational rate of the pulsar itself, meaning that it falls outside the typical observational energy ranges. Moreover, modern ideas suggest a co-rotating, plasma-filled (close to force-free) magnetosphere. If magnetic dipole radiation occurs in pulsar magnetospheres, it should be reprocessed to accelerate particles to higher energies, and they should in turn radiate emission at frequencies accessible to multi-wavelength satellites and telescopes.} from the WD, establishing AR Sco as the first known\footnote{We note, however, that there have been previous speculations of both X-ray and radio burst data invoking the idea of a WD pulsar; see Section \ref{sec:5}.} WD pulsar, analogous to rotation-powered neutron star pulsars. Recently, a claim was made for a second such system, based on 0.8 magnitude coherent variations in the optical, with a periodicity of 12.37 minutes \citep{Kato2021}. 

The WD in AR Sco is believed to be highly magnetized, with a polar surface field of $B_{\rm p}\sim8\times10^8$~G (inferred by equating the spin-down loss rate to the magnetic dipole loss rate) and its optical emission is strongly linearly polarized (up to$\sim40\%;$ \citealt{Buckley2017}). In the radio, the degree of linear polarization is much lower, $< 1\%$, possibly due to this emission coming from non-relativistic plasma \citep{Stanway2018}. Extensive optical observations indicated that the linear flux, circular flux, and polarization position angle (PPA) are coupled to the spin period $P$ of the WD and not the beat period \citep{Potter2018b}. 
The spin phase-folded PPA data from these observations clearly exhibited periodic emission vs. WD spin phase $(\phi_{\rm s})$ with a $180^{\circ}$ sweep in PPA. This was consistent with the light curve from \cite{Buckley2017}, where a double-peaked structure manifested with an intense first peak and a dimmer second peak (these peaks are separated by $180^{\circ}$in spin phase). These facts imply that the WD may be an orthogonal rotator if the emission originates close to its polar caps \citep{Geng2016,Buckley2017}. This is also confirmed by more recent observations by \citet{Garnavich2020} that infer two hotspots yielding very symmetric spin modulation of the FUV flux, and implying that the hotspots are very close to the WD spin equator, thus lying near the WD magnetic poles.
Moreover, \citet{Marsh2016} found that the companion has an irradiated side facing the WD, contributing to the observed sinusoidal radial velocity profile. This suggests that the two stars are tidally locked \citep{Buckley2017, Takata2017, Potter2018}. 

Several different models have been proposed (see Section~\ref{sec:5}) to account for the observed non-thermal radiation, placing the emission at different spatial locale and invoking different radiation scenarios. In one of the first models, \cite{Geng2016} noted that the Goldreich-Julian (GJ) charge number density \citep{Goldreich1969} of the WD is much lower than that required by the observed SR spectrum and thus argued that the emission should originate closer to the companion (they suggest at an intrabinary shock caused by the interacting stellar winds; \citealt{Marsh2016,Geng2016}), where the particle number density is presumably much higher than the GJ density. However, the majority of models speculate that relativistic electrons are being injected along the magnetic field lines (toward the WD surface), where they are accelerated and trapped in the WD magnetosphere \citep{Takata2017,Buckley2017,Lyutikov2020}. These may radiate from near the magnetic poles, with the emission from downward-moving particles being directed toward the WD. There thus seems to be a majority consensus that the pulsed non-thermal emission is due to magnetospheric SR from the WD. This is supported by the geometric model of \cite{Potter2018b} that can reproduce the observed polarization signatures if the emission location is taken to be near the WD poles. 

In our previous work \citep{DuPlessis2019}, we modelled the linear polarization signature with the rotating vector model (RVM; \citealt{Radhakrishnan}), defining $\alpha$ as the magnetic inclination angle of the magnetic dipole moment $\bm{\mu}$ and $\zeta$ as the observer angle (observer's line of sight), both measured with respect to the WD's rotation axis. We constrained $\alpha$, $\zeta$ and the WD mass. Our solution for $\alpha$ using the RVM supports the conjecture of an orthogonally-rotating WD. We demonstrated that the RVM provides an excellent fit to the PPA curve, thereby favouring the magnetospheric scenario, if particles can sustain small pitch angles.  

In this paper we aim to show that the standard RVM describes the optical orbital phase-resolved polarimetric observations reasonably well, reaffirming the conclusions our previous work. We also report that the variation in $\alpha$ and $\zeta$ over the orbital period is a robust recurrent signature and we do not believe that it is the fitting procedure that is compensating for a simplistic model. 
  
The structure of this paper is as follows. In Section~\ref{sec:3}, our implementation of the RVM for the phase-resolved data will be discussed and the results follow in Section~\ref{sec:4}. In Section~\ref{sec:5}, we discuss existing models and current observations. Finally, we propose possible emission scenarios for AR Sco in  Section~\ref{sec:6}.

\section{Method}\label{sec:3}
We apply the RVM to the extensive optical observations of \cite{Potter2018} in order to probe $\alpha$ and $\zeta$ of the WD as a function of orbital period in a phase-resolved analysis. The new data include observations performed on 27-28 March 2018 and 14, 25, 26, and 27 May 2018 with the high-speed photo-polarimeter on the 1.9~m optical telescope located at Sutherland. The observations have been converted to Barycentric coordinates; for more details, see \cite{Potter2018}. \cite{Gaibor2020} recently computed the effect of the proper motion of the system on the inferred spin down of both the orbital period and beat period. This effect was found to be four orders of magnitude smaller than the first $700$ times smaller than the second, making the Shklovskii effect  \citep{Shklovskii1970} negligible for the system. The light crossing timescale ($a/c \approx 2.5$~s) is also small compared to the orbital period and WD spin period, so that  relativistic corrections may be neglected.     

The data included the Barycentric Julian Day (BJD), Stokes $Q$, and Stokes $U$ parameters with their errors, orbital phase $\phi_{\rm b}$, spin phase $\phi_{\rm s}$, and beat phase. Each file had an accompanying file with the total intensity measurement and the total observational time amounted to $\sim30$ hours. As per standard convention, $\phi_{\rm b}=0$ is set at the inferior conjunction of the companion.  

We binned the data into 50 spin-phase bins and 100 orbital phase bins. 
We next propagated the $Q$ and the $U$ Stokes parameter errors when calculating the PPA using $\psi = 0.5\arctan\left(U/Q\right)$, as follows
\begin{equation} \label{err_prop}
\delta \psi^{2} = \left(\frac{\partial \psi}{\partial U}\right)^{2}\delta U^{2} +  \left(\frac{\partial \psi}{\partial Q}\right)^{2}\delta Q^{2} + 2\frac{\partial \psi}{\partial U}\frac{\partial \psi}{\partial Q}\delta_{\rm UQ},
\end{equation}
where $\delta_{\rm UQ}$ is the covariance of $U$ and $Q$. By substituting 
\begin{equation}
\begin{split}
\frac{\partial \psi}{\partial U} &= \frac{1}{2}\frac{1}{1 + \left(U/Q\right)^{2}}\frac{1}{Q}, \\
\frac{\partial \psi}{\partial Q} &= -\frac{1}{2}\frac{1}{1 + \left(U/Q\right)^{2}}\frac{U}{Q^{2}} \\
\end{split}
\end{equation}
into Equation~(\ref{err_prop}) and after some algebra, finally yields
\begin{equation} \label{PPA_err_prop}
\delta \psi^{2} = \left(\frac{1/2}{1 + \left(U/Q\right)^{2}}\right)^{2}\left(\frac{U}{Q}\right)^{2}\left[\left(\frac{\delta U}{U}\right)^{2} + \left(\frac{\delta Q}{Q}\right)^{2} -\frac{2\delta_{\rm UQ}}{UQ}\right].
\end{equation}
The last term in the above equation only appears if $U$ and $Q$ are correlated, in which case the error on $\psi$ will decrease. The value of $\delta_{\rm UQ}$ was difficult to estimate; since the fractional errors $\delta U/U$ and $\delta Q/Q$ may be small in some cases, even a small value of $\delta_{UQ}$ may cause the last term in Equation~(\ref{PPA_err_prop}) to become dominant, yielding an imaginary error $\delta\psi$. In the light of these problems, we conservatively overestimated the errors by setting $\delta_{\rm UQ} = 0$, meaning we assumed that $U$ and $Q$ are uncorrelated and the last term is zero. 

Next, we implemented a similar method as discussed in \citet{DuPlessis2019} to fix the convention issues relating to $\psi$ and to generate a continuous PPA curve. We then double-checked the data manually to ensure that we had smooth, continuous PPA curves as a function of binary phase. 
We then fit the RVM
\begin{equation}
\tan(\psi-\psi_{0})=\frac{\sin\alpha\sin(\phi_{\rm s}-\phi_{0})}{\sin\zeta\cos\alpha-\cos\zeta\sin\alpha\cos(\phi_{\rm s}-\phi_{0})}
\end{equation}
to the PPA data and applied the emcee Markov Chain Monte Carlo code \citep{Foreman2013} to find the best-fit parameters, yielding $\alpha$ and $\zeta$ for each $\phi_{\rm b}$ bin, using 32 walkers, 25,000 steps and a burn-in value of 8,000. More dense sampling with more walkers or steps did not alter our best-fit results.

To test the robustness of our results, we explored different binning combinations of the the data sets, as well as fitting the RVM as a time series instead of folding the data using the ephemeris. The different binning combinations are shown in the next Section, indicating that we found similar best-fit results. Second, we found the best fits for $\alpha$ and $\zeta$ by minimizing a $\chi^2$ statistic and found that the results are generally consistent with those derived from the MCMC approach, even if we change the initial-guess values. If the PPA fit to the data is particularly bad, the $\chi^2$ can end up giving best-fit values for $\alpha$ and $\zeta$ associated with local minima, which is not desirable.

\section{Results}\label{sec:4}
\subsection{Individual PPA vs.\ $\phi_{\rm s}$ Fits}
\begin{figure}
\centering
\includegraphics[width=20pc]{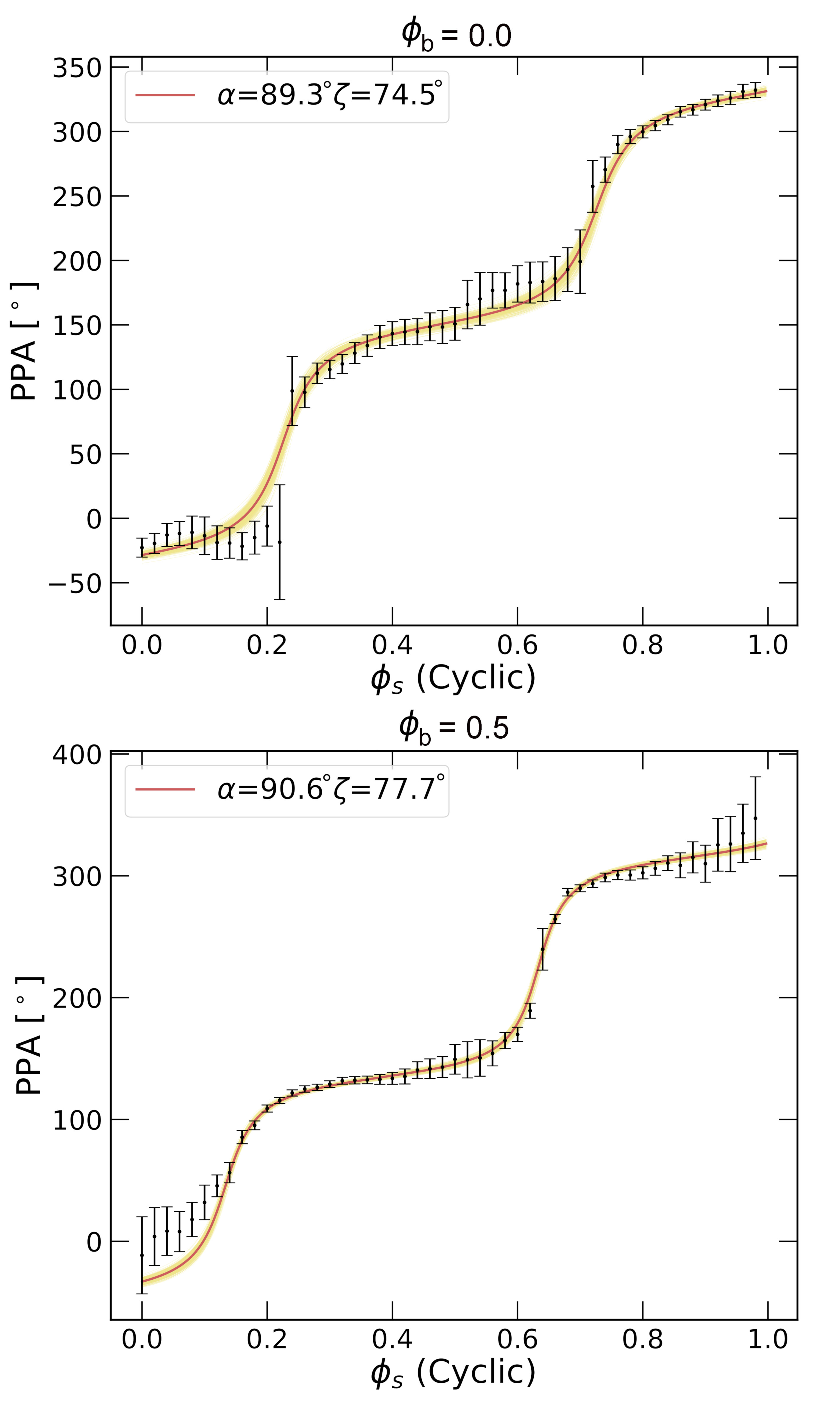}
\caption{The best-fit results to the PPA data from \protect\cite{Potter2018b} at different values of $\phi_{\rm b}$ (as indicated by labels for each panel), where the first panel is at $\phi_{\rm b} = 0$ and the second at $0.5$. The solid red line indicates the best RVM fit, with the yellow being ensemble fits within the $16^{\rm th}$ and $84^{\rm th}$ percentile posteriors.}
\label{Phase_fits1}
\end{figure}
\begin{figure}
\centering
\includegraphics[width=20pc]{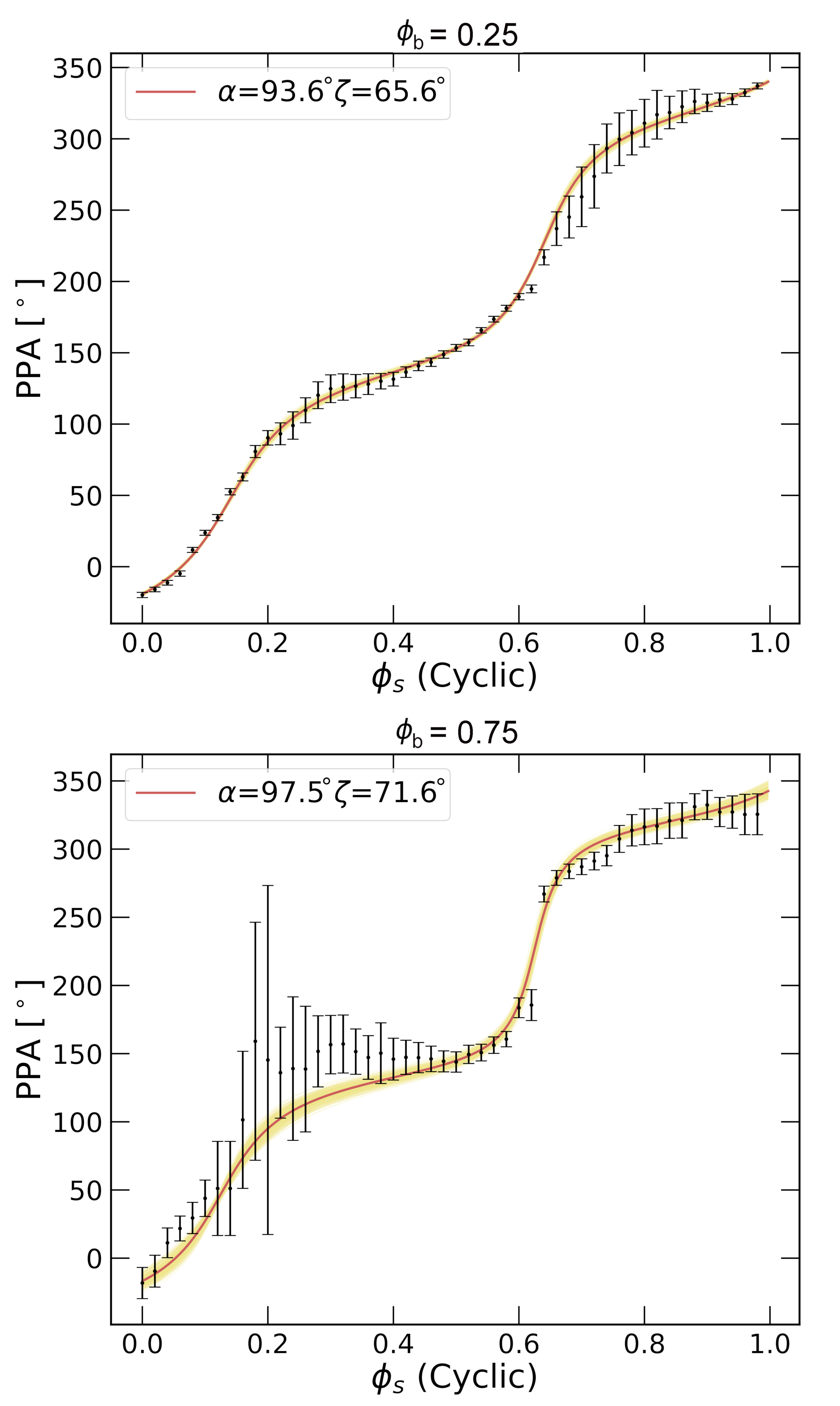}
\caption{The best-fit results to the PPA data from \protect\cite{Potter2018b} at $\phi_{\rm b}=0.25$ (top panel) and $\phi_{\rm b}=0.75$ (second panel). As before, the solid red line indicates the best fit with the yellow being ensemble fits within the $16^{\rm th}$ and $84^{\rm th}$ percentile parameter errors.}
\label{Phase_fits2}
\end{figure}

Previously \citep{DuPlessis2019}, we fit the PPA data from \cite{Buckley2017}, which were averaged over a large range of $\phi_{\rm b}$. Here, we fit the RVM to phase-resolved PPA data from \cite{Potter2018b} to investigate the evolution of the geometric parameters ($\alpha$ and $\zeta$) we previously constrained over a range binary phases. The panels in Figures~\ref{Phase_fits1} and~\ref{Phase_fits2} show that the RVM describes the broad structure of the observed data at the different $\phi_{\rm b}$ quite well. Interestingly, there is a slight change in PPA structure with $\phi_{\rm b}$ that the model is able to capture, but this implies a modulation of $\alpha$ and $\zeta$ with $\phi_{\rm b}$.

\subsection{Obtaining $\alpha$ and $\zeta$ vs.\ $\phi_{\rm b}$}

\begin{figure}
\centering
\includegraphics[width=20pc]{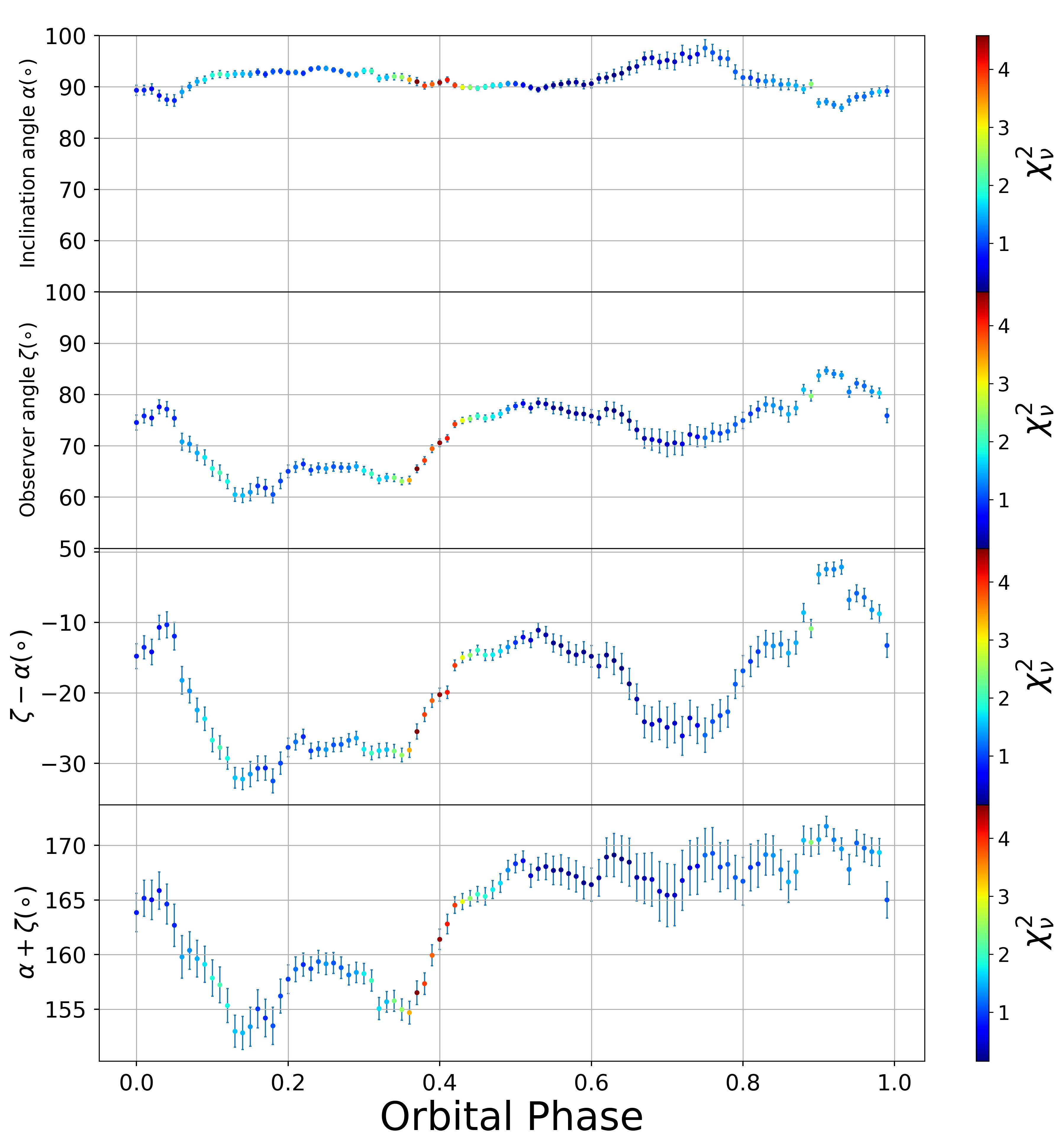}
\caption{The first panel shows the best-fit $\alpha$ values at each $\phi_{\rm b}$, obtained with the MCMC code using the data from \protect\cite{Potter2018b}. The colour bar indicates the reduced $\chi^2$ value for each such PPA fit at fixed $\phi_{\rm b}$ (see Figure~\ref{Phase_fits1}). The second panel shows the accompanying $\zeta$ vs.\ $\phi_{\rm b}$ best-fit results obtained from the MCMC. In the third panel, $\beta=\zeta-\alpha$ is plotted while the fourth shows $\alpha+\zeta$ vs.\ $\phi_{\rm b}$.}
\label{Orb_fit}
\end{figure}

By fitting the RVM to the PPA vs.\ $\phi_{\rm s}$ data at each $\phi_{\rm b}$, one can probe $\alpha(\phi_{\rm b})$ and $\zeta(\phi_{\rm b})$. In the first two panels of Figure~\ref{Orb_fit}, such phase-resolved best-fit values for $\alpha$ and $\zeta$ are shown. The variation in $\alpha$ seems to be semi-stable, since the minimum to maximum variation is only about $\sim 10^{\circ}$. The variation in $\zeta$ is much larger at about $\sim 30^{\circ}$, intriguingly with four bumps visible close to $\phi_{\rm b}=0.0, 0.25, 0.5$. The bump close to $\phi_{\rm b}=0.75$ is not as clearly defined as the others, but note that this may be associated with the dip in PPA vs.\ orbital and $\phi_{\rm s}$ around and slightly after $\phi_{\rm b}=0.75$, as seen in the last row of Figure~2 of
\citet{Potter2018b}. 
In the third panel, these dips are also discernible in the ``impact angle'' $\zeta - \alpha$ vs.\ $\phi_{\rm b}$. 
The last panel of Figure~\ref{Orb_fit} indicates that there is clear variation in $\alpha + \zeta$ for the first part of the orbital period, indicating that $\alpha$ and $\zeta$ are not purely anti-correlated. This suggests that the model is not trying to compensate for not being able to fit the data by making the parameters anti-correlating. In fact, the RVM is always able to describe the broad structure of the PPA, thus, the variation of $\alpha$ and $\zeta$ with $\phi_{\rm b}$ seems intrinsic to the emission geometry. Relatively large reduced $\chi^2$ values are found at $0.4-0.5$ in orbital phase (red dots). This is due to the small formal errors in the PPA at these points in $\phi_{\rm b}$, given that the linear flux is at its maximum there. 

\subsection{Robustness of Variation of $\alpha$ and $\zeta$ with $\phi_{\rm b}$}
\begin{figure}
\centering
\includegraphics[width=20pc]{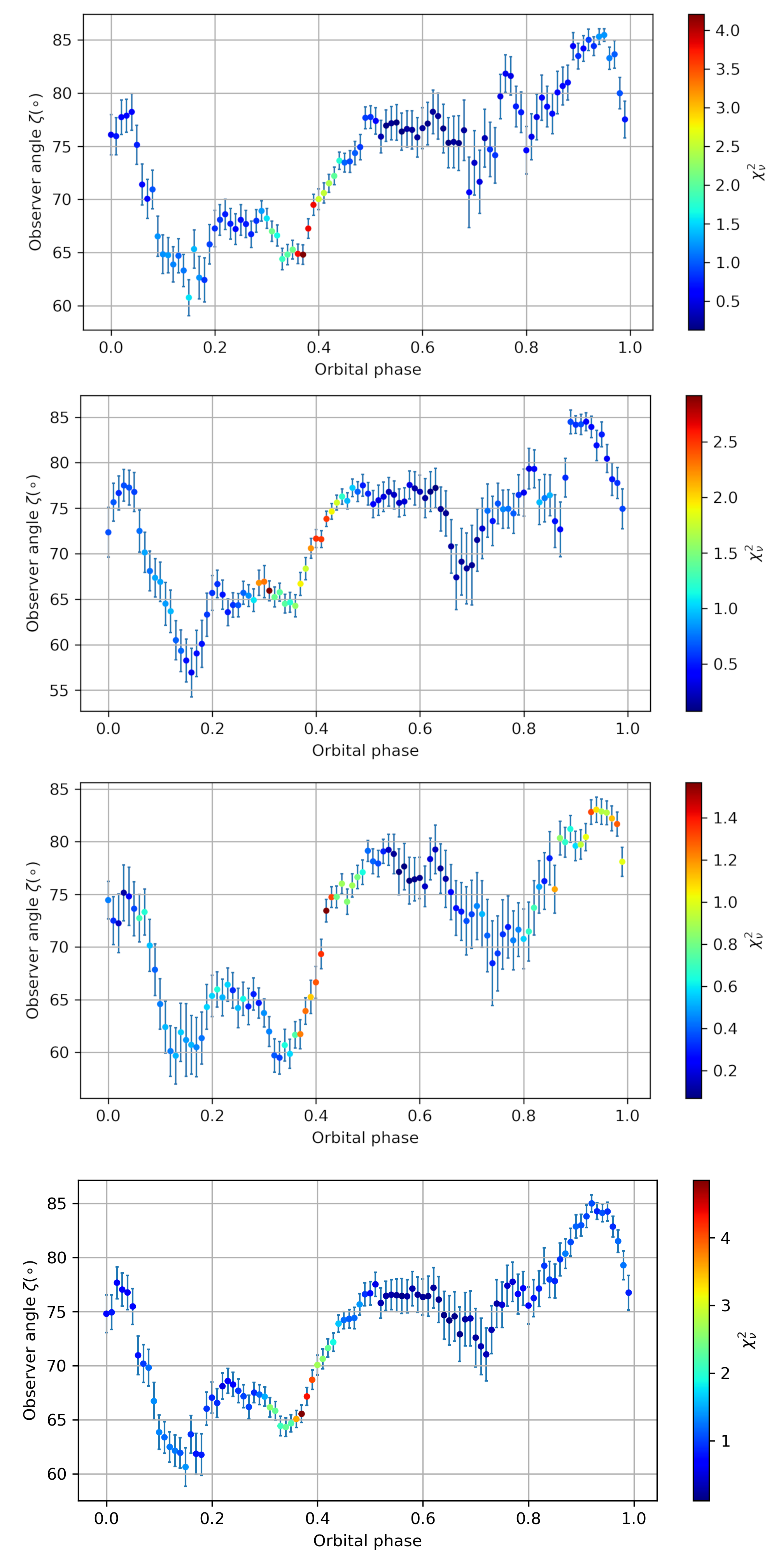}
\caption{Best-fit results for the observer angle at different $\phi_{\rm b}$ for various epoch-binning combinations. The first panel includes the data for 25, 26 and 27 May, the second panel includes those from 26 May and 27 March, the third panel includes data from 27 May and 28 March, and the fourth panel includes data from 14, 25, 26 and 27 May. All these data epochs were taken in 2018.}
\label{Bin_fits1}
\end{figure}
\begin{figure}
\centering
\includegraphics[width=20pc]{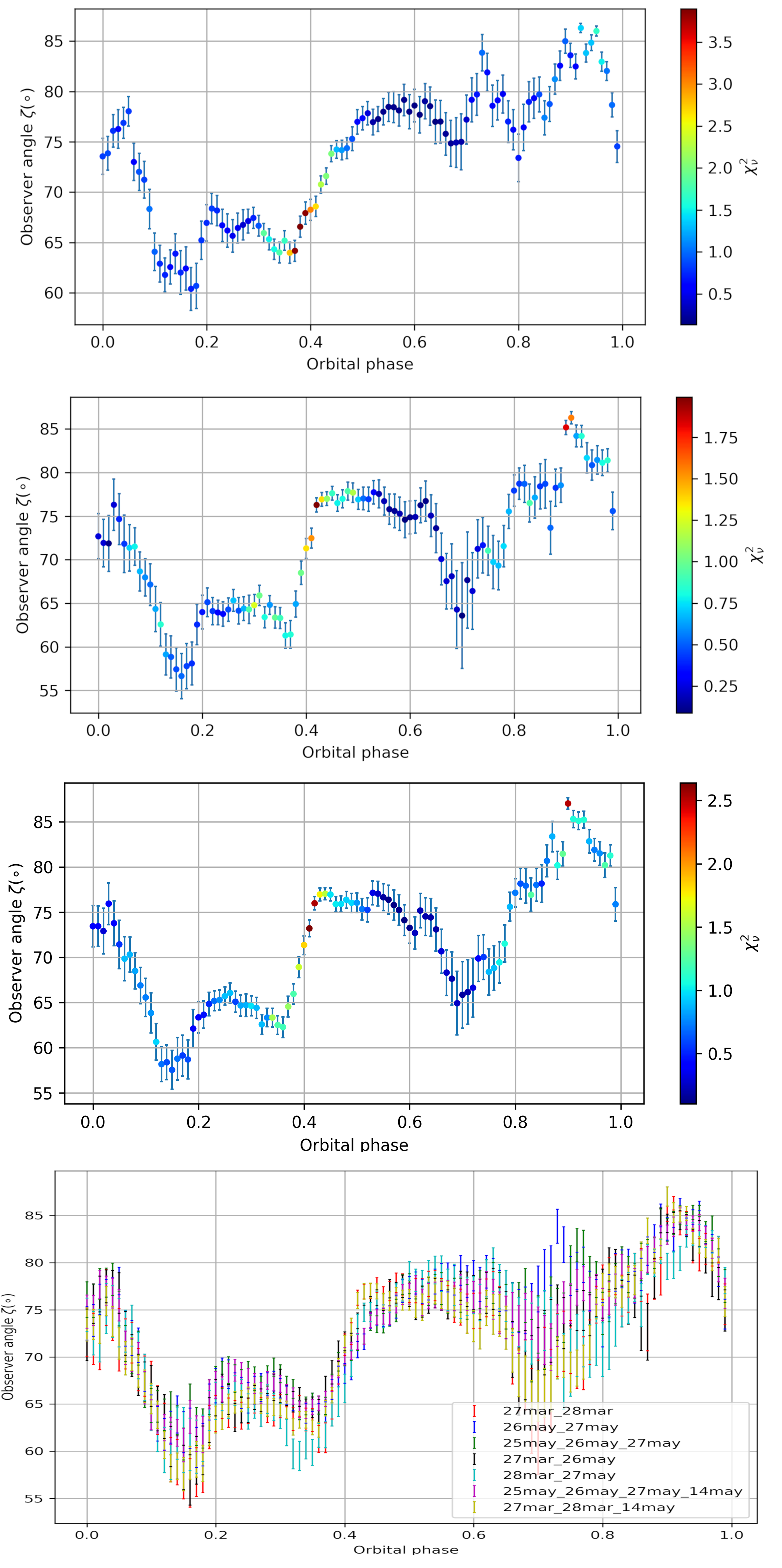}
\caption{Similar to Figure~\ref{Bin_fits1}
, but for data for 26 and 27 May (first panel), 27 and 28 March (second panel), 27, 28 March and 14 May (third panel), and the final panel contains results for all seven binning combinations. All these data epochs were taken in 2018.}
\label{Bin_fits2}
\end{figure}
In order to ascertain the robustness of the variation of $\alpha$ and $\zeta$ with $\phi_{\rm b}$, and whether this variation is stable, we investigated different binning combinations and different data epochs.

The panels of Figures~\ref{Bin_fits1} and~\ref{Bin_fits2} suggest that the structure over $\phi_{\rm b}$ of $\zeta$ in Figure \ref{Orb_fit} is independent of the epoch-binning choices, with the only discrepancy found at phase $0.7-0.85$ where there is low linear flux, causing large formal errors. All the panels of Figures~\ref{Bin_fits1} and~\ref{Bin_fits2}, with the exception of panel 3 of both, show a peak at $\phi_{\rm b}=0.75$; the absence of this peak is also visible in the last panel of Figure~\ref{Bin_fits2} where all the epochs were binned together. It would seem that the peak at $\phi_{\rm b}=0.75$ is more visible for the May observations; this could be due to the fact that each March observation was about $60-80\%$  shorter than the individual May observations. 

The plots of Figure~\ref{Bin_fits1} and~\ref{Bin_fits2} were similarly done for the magnetic inclination angle $\alpha$ as well, but were not included here due to the small variations in $\alpha$ vs.\ $\phi_{\rm b}$. The individual data sets were not used on their own due to gaps resulting from calibration observations. 

\begin{figure}
\centering
\includegraphics[width=20pc]{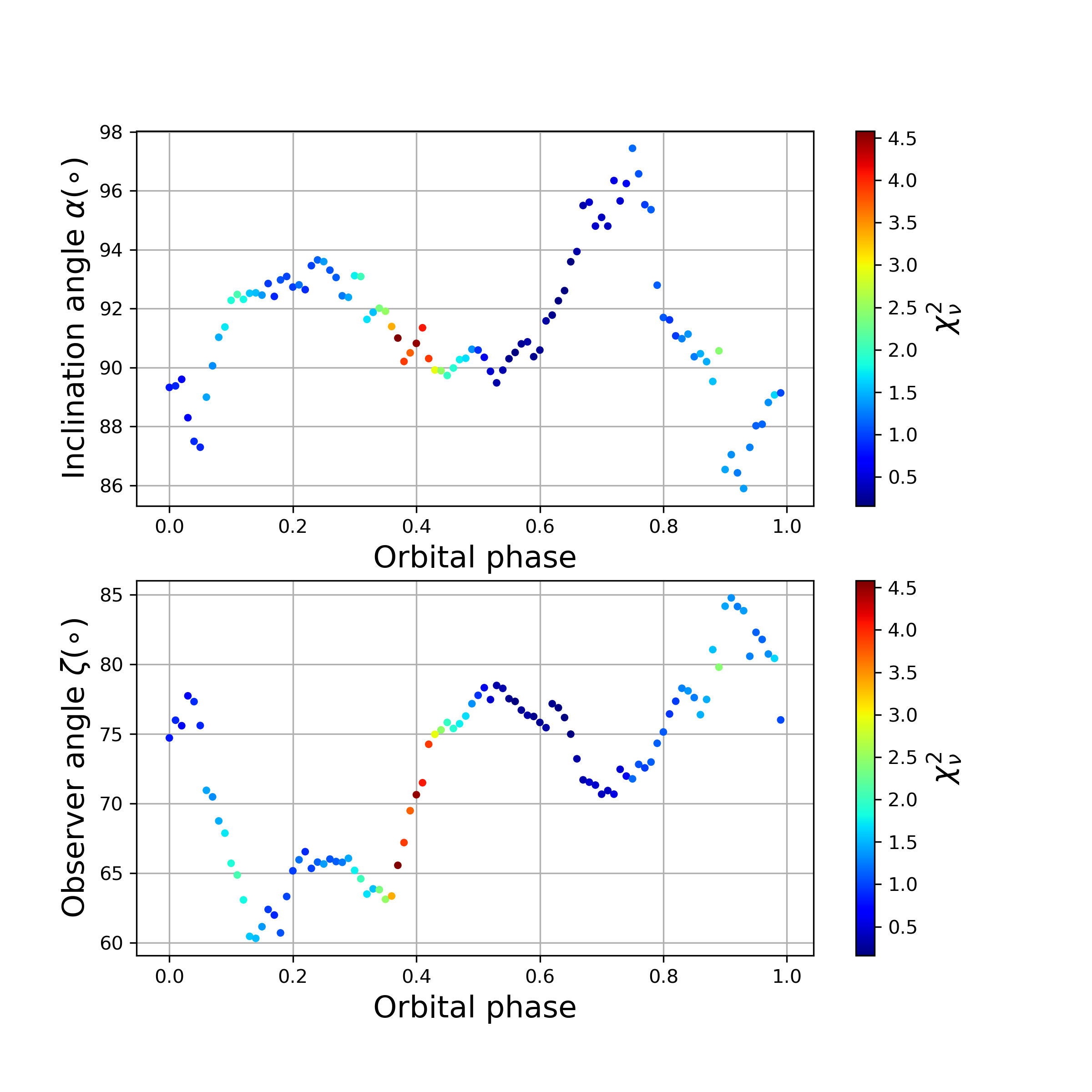}
\caption{The minimized $\chi^2$ fit for the RVM at the different $\phi_{\rm b}$.}
\label{Orb_fit_chi}
\end{figure}

In Figure~\ref{Orb_fit_chi} the minimized $\chi^2$ results are shown for $\alpha$ and $\zeta$ vs.\ $\phi_{\rm b}$, as a basis of comparison with our previous results that were obtained using an MCMC (log-likelihood) approach. These plots do not contain errors, since the $\chi^2$ method does not yield errors for the best-fit values, but only indicates a goodness-of-fit value (this was only done to compare best-fit values, without exploring the phase space using the $\chi^2$ statistic). Comparing these results with Figure~\ref{Orb_fit} gives us more confidence in our results, since the two methods yield very similar posterior best-fit values. 

\subsection{Spin and Beat Light Curves vs. $\phi_{b}$}
To investigate how the emission components respectively coupled to the beat and spin periods contribute to the linear flux over $\phi_{b}$, we plot the light curves at different orbital phases for the spin-folded linear flux and beat-folded linear flux. We include the spin-folded PPA with our RVM best fit, as well an inset depicting the binary configuration at the indicated orbital phase.

\begin{figure}
\centering
\includegraphics[width=20pc]{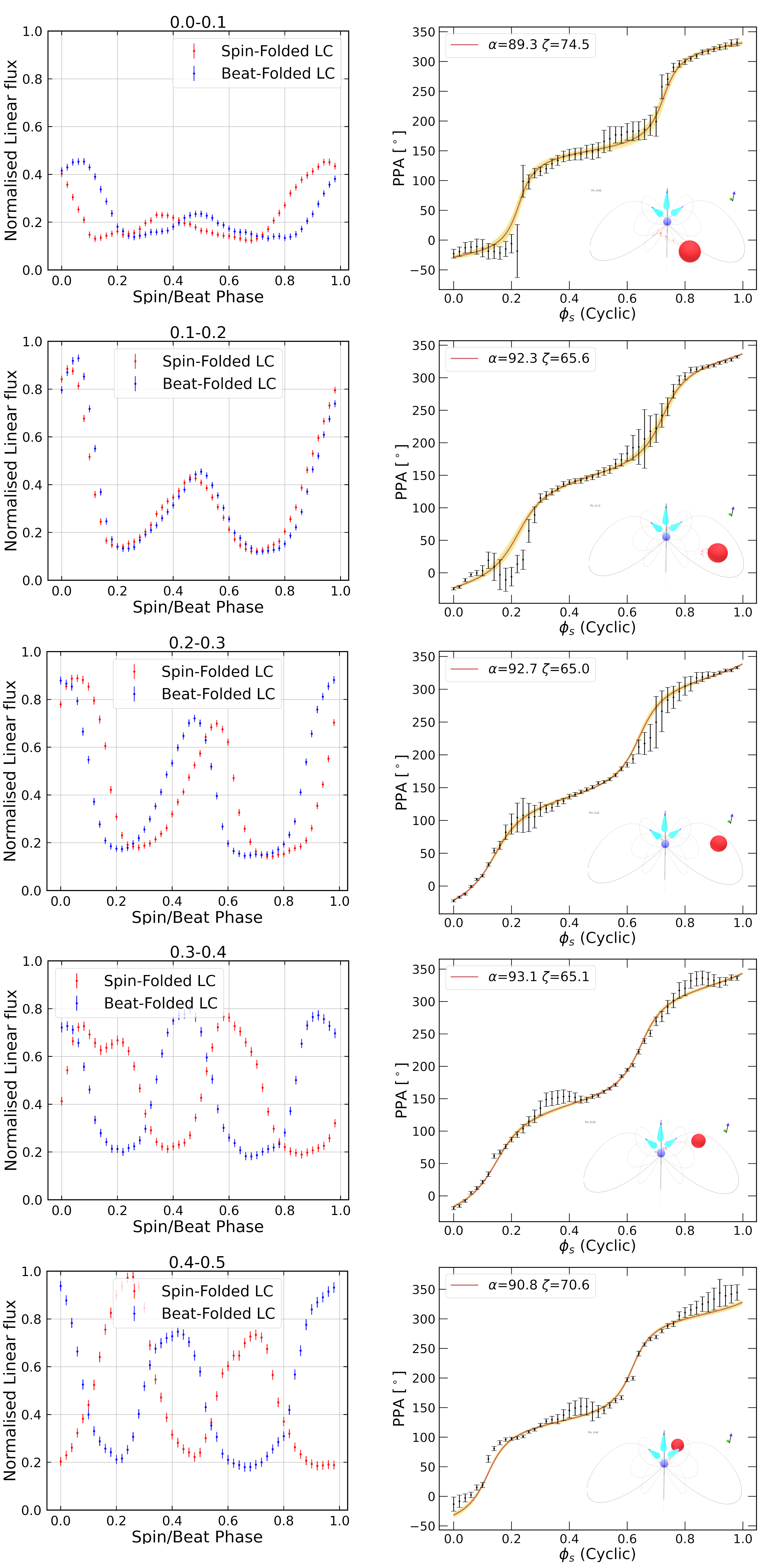}
\caption{Spin/Beat-folded light curve and PPA data integrated over $10\%$ of the orbit. The first column contains the spin-folded light curves of the normalized linear flux (i.e., the maximum flux over the orbital period is unity) in red and the beat-folded light curves of the normalised linear flux in blue. The second column shows the spin-folded PPA. Both columns range from $0.0-0.5$ in orbital phase.}
\label{Orb_spin_LC_PPA0.5}
\end{figure}

\begin{figure}
\centering
\includegraphics[width=20pc]{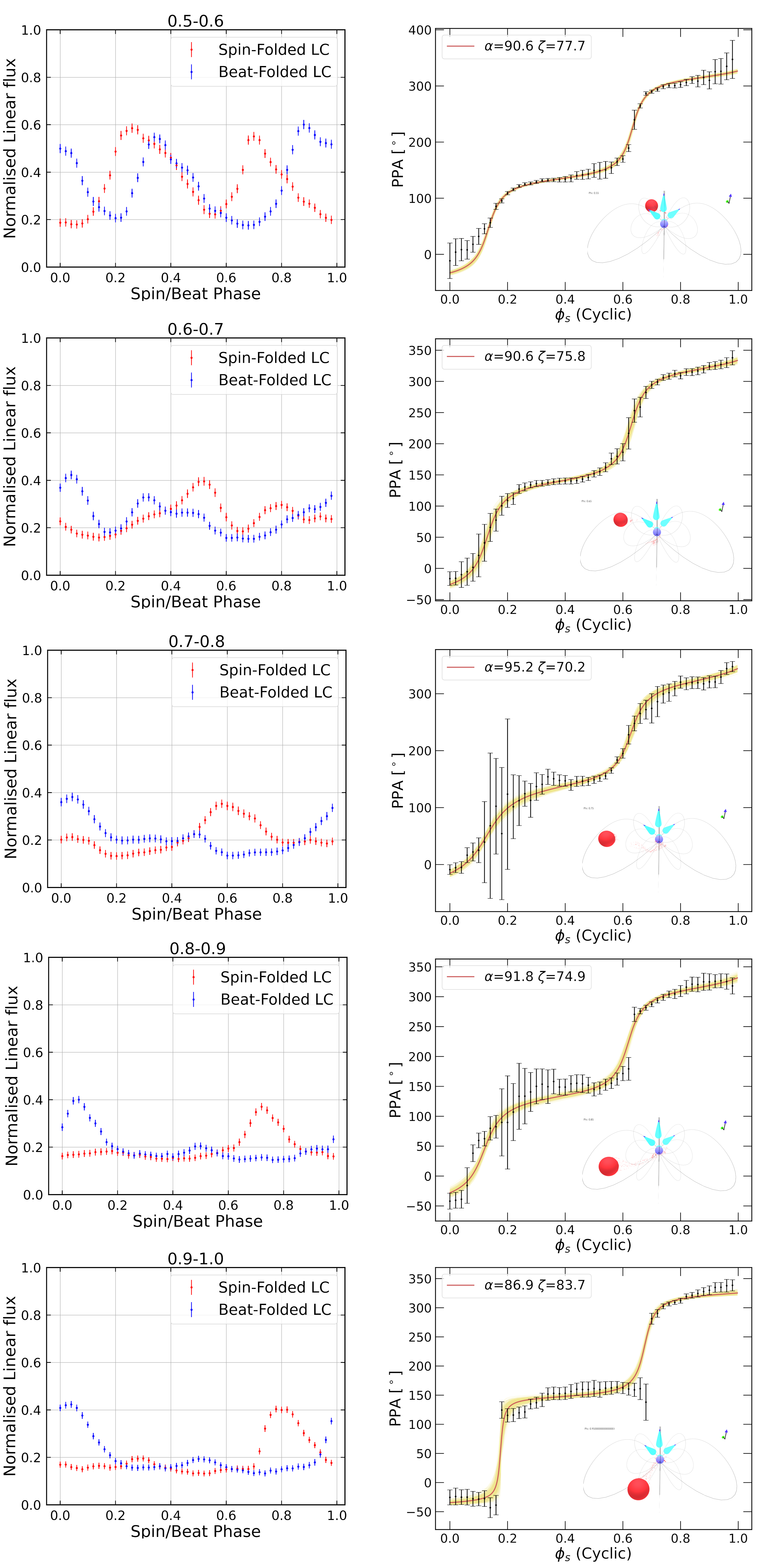}
\caption{
Same as Figure~\ref{Orb_spin_LC_PPA0.5}, but for $\phi_{\rm b}=0.5-1.0.$
}
\label{Orb_spin_LC_PPA1.0}
\end{figure}
The first column of Figures \ref{Orb_spin_LC_PPA0.5} and \ref{Orb_spin_LC_PPA1.0} show a clear evolution of the spin- and beat-folded light curves over the orbital period. The beat-folded light curves appear more stable over the full $\phi_{b}$, while the spin-folded light curves generally move rightward. At these orbital phases, as well as the first panel of Figure~\ref{Orb_spin_LC_PPA1.0}, the beat-folded light curves start to move leftward and the spin-folded light curves seem to be more stable. This suggests that at this region in orbital phase, the spin periodicity is stronger, while the beat periodicity seems more significant at other phases. Interestingly, the linear flux is also relatively larger at these phases compared to the rest of the orbital phases, corresponding to the maximum linear flux in the pulse profile plots from \citet{Potter2018}. 
In Figure \ref{Orb_spin_LC_PPA0.5}, in the first two rows of column two, we see that segments of the PPA are discontinuous; this is also visible in the last two rows of Figure~\ref{Orb_spin_LC_PPA1.0}, namely close to $\phi_{\rm b}=0.0$ where the companion is between the observer and the WD. 

\begin{figure}
\centering
\includegraphics[width=20pc]{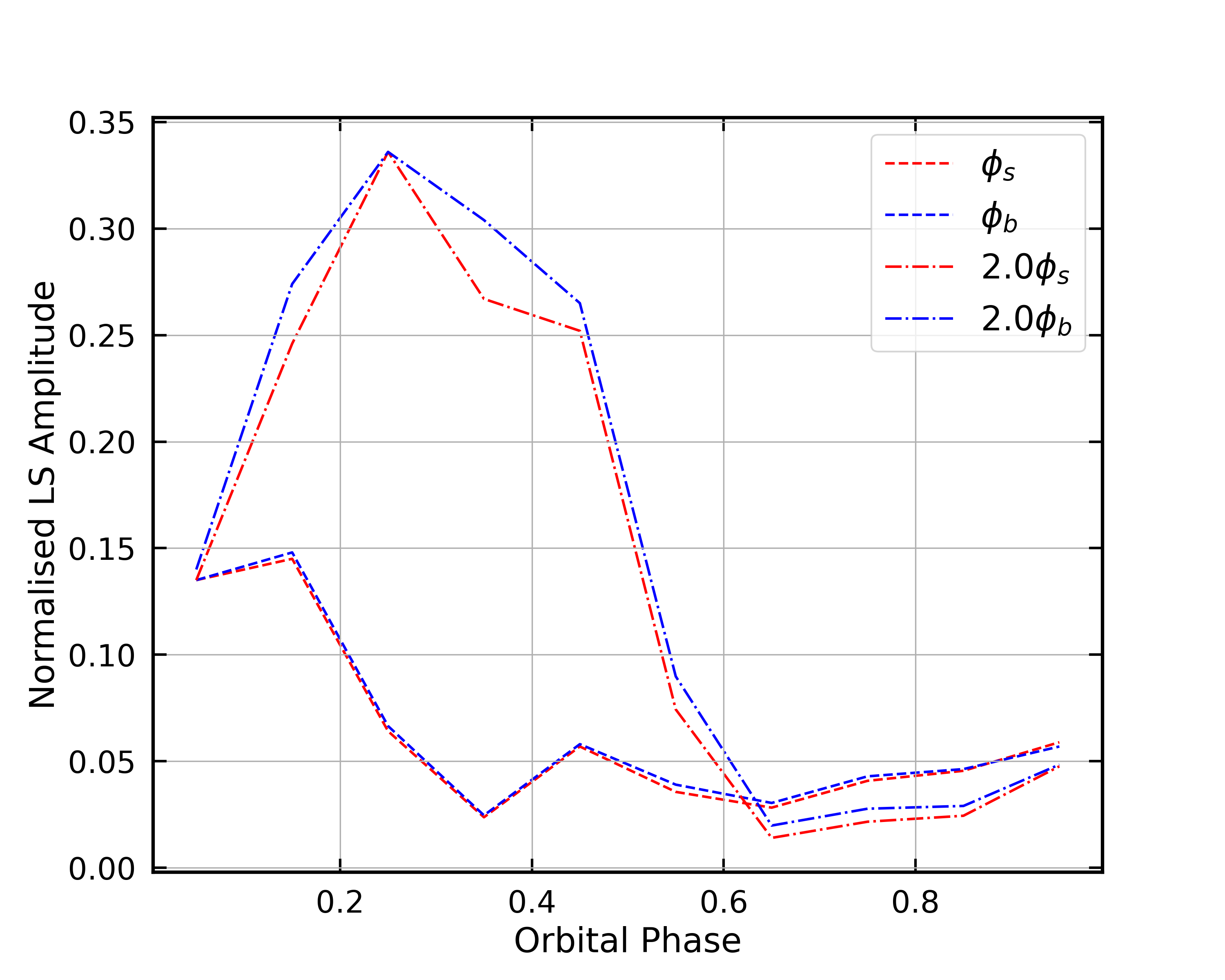}
\caption{Lomb-Scargle amplitudes of the first and second harmonics of the spin and beat frequencies.}
\label{Lomb_amp}
\end{figure}

To further probe which periodicity dominates at the different orbital phases, we performed a Lomb-Scargle analysis \citep{VanderPlas2018} of the ten segments used for the light curves in Figures~\ref{Orb_spin_LC_PPA0.5} and \ref{Orb_spin_LC_PPA1.0} (for orbital phase bins of 0.1). These results are shown in Figures~\ref{FFT_orbital1} and \ref{FFT_orbital1}, where each panel is at a different orbital phase. Several vertical lines are used to indicate first, second and third harmonics and side-band frequencies for the spin and beat periods, as indicated in the legend. To save space, we concatenated zoomed-in panels to indicate non-zero amplitudes for each of these harmonics. The orbital phase range is indicated as a title, for each panel.
For clarity, we included Figure~\ref{Lomb_amp} to show the first and second harmonic normalized amplitudes of the spin and beat frequency vs.\ $\phi_{b}$.

\begin{figure}
\centering
\includegraphics[width=20pc]{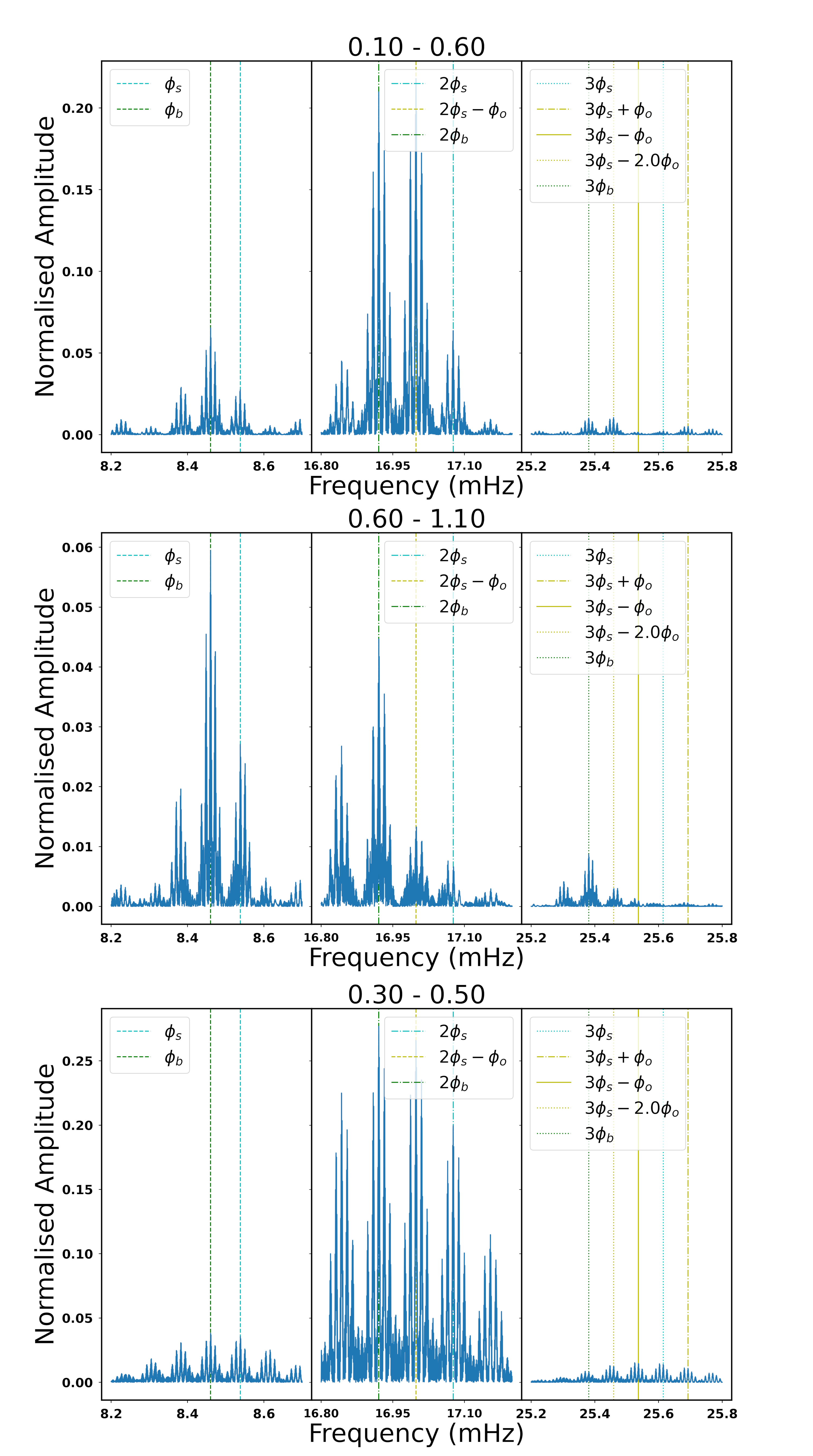}
\caption{Lomb-Scargle power spectrum for different segments in orbital phase. In each panel, the beat frequency and its harmonics are indicated by green lines, the spin and its harmonics by cyan lines, and other notable side-band frequencies by yellow lines.}
\label{FFT_orbital_2bins}
\end{figure}

Interestingly between orbital phases $0.1$ to $0.6$, the second harmonics of the spin and beat frequency have a much higher normalized amplitude than the first harmonics. During orbital phase $0.6$ to $1.1$, we see that the first harmonics' normalized amplitude is slightly higher than the second harmonic's, but both are at a significantly lower level than the second harmonic's amplitude in the first segment of the orbital phase. Figure~\ref{Lomb_amp} also shows that the beat and spin normalized amplitudes are quite similar, thus in Figure~\ref{FFT_orbital_2bins} we used larger segments of the orbital phase: $0.1 - 0.6$, $0.6 - 1.1$, and $0.3 - 0.5$, with the first harmonics, second harmonics, third harmonics and some side-band frequencies shown. In the $0.6 -0.1$ range, we see once again that the second harmonics are at a much higher amplitude than the first harmonics; additionally the amplitudes of the harmonics of the beat frequency are slightly larger than the spin frequency harmonics. In the region $0.6 - 1.1$, the first harmonics are at a slightly larger amplitude than the second harmonics, with the beat-frequency harmonics being much larger than the spin-frequency harmonics. In the region $0.3-0.5$, the linear flux is maximal; \citet{Potter2018} suggest that this is where particles are injected into the WD magnetosphere. We see substantially larger second harmonics in this range compared to first harmonics, with equal beat and spin-frequency first harmonics, with a slightly larger second harmonic for the beat frequency compared to the spin frequency.

\begin{figure}
\centering
\includegraphics[width=16pc]{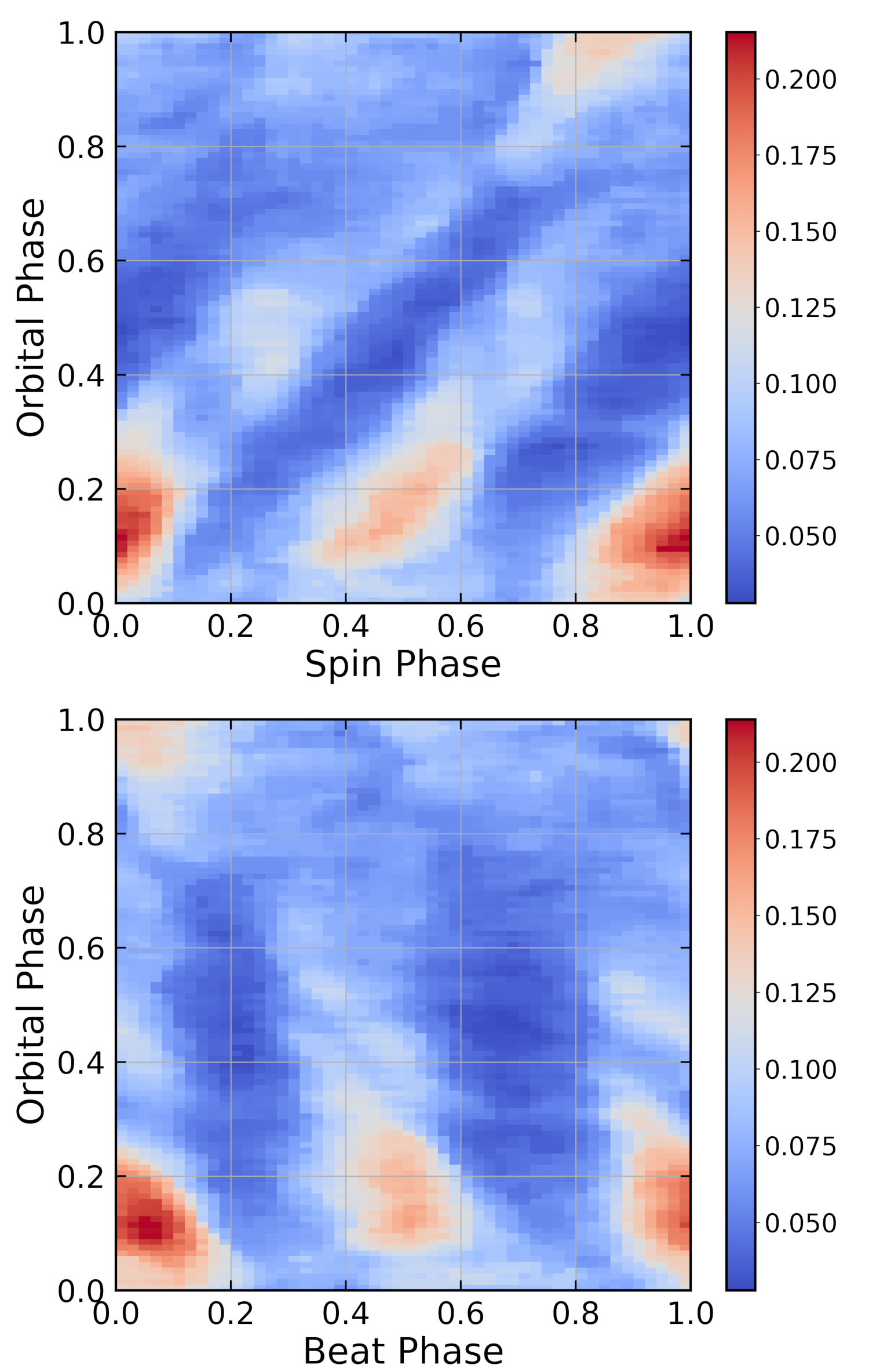}
\caption{Phase plots of the degree of polarisation for the data from \citet{Potter2018} plotted using a colour scale. The first panel shows the beat-folded degree of polarisation over the orbital phase and the second shows the spin-folded data.}
\label{Deg_Lin}
\end{figure}

In Figure~\ref{Deg_Lin} we plot the beat-folded and the spin-folded degree of polarisation as obtained from the observational data. We see clear evolution of this quantity over the orbital period, with the maximum occurring just before 0.2 in $\phi_{\rm b}$. Both panels show a higher degree of polarisation at $0.85-0.15$ and $0.4-0.6$ in both $\phi_{\rm s}$ and beat phase. In the second panel, the regions of low degree of polarisation (dark blue lines) seem to be coupled to the beat period, since these lines are more vertical than in the first panel. For the regions of higher degree of linear polarisation,  both panels seem to have slanted elements, suggesting a mixture of emission components. For a constant spectral shape of synchrotron-emitting particles, one would expect a fixed value of the degree of polarisation. Thus, an evolution in polarisation degree may either be due to orbitally-dependent depolarisation, or result from an evolving index of the non-thermal electrons (or a mixture of both effects). In the latter case, such evolution may hint at the acceleration or cooling of particles at different orbital phases, barring the first effect.

\section{Discussion}\label{sec:5}
The first mention of a ``WD pulsar'' producing high-energy emission via a pulsar-like mechanism was probably made by \cite{Usov1988}. Other ideas invoking WD pulsars were proposed by \citet{Paczynski1990,Usov1993,Hulleman2000} to account for anomalous X-ray pulsars (magnetars). 
\cite{Zhang2005} proposed that the 5 radio outbursts of the transient source GCRT J1745$-$3009 could be due to a WD pulsar, due to the pulsar-like periodic nature of the emission.

Below, we briefly discuss selected modeling attempts, linking these to available observations. We categorised these models into three groups, namely: interacting magnetic winds, WD magnetosphere and intermediate polar models. 

\subsection{Magnetic-Wind Interaction Models}
These models place the origin of the emission at the interacting magnetized winds of the two binary members, where particle densities are higher and particles could be accelerated to higher energies. However, that some of these models may have difficulty to reproduce the highly polarized optical emission seen from the system.

One of the first models proposed for AR Sco was that by \cite{Geng2016}, where they assume a magnetic dipole field geometry and estimate the field strength at the interface of interacting stellar winds or close to the surface of the companion to be $B\sim 1,860 \, \rm{G}$. They take the non-thermal emission to be SR, the spectrum of which is described by a broken power law, and derive a critical Lorentz factor $\gamma_{\rm c}=73$ using the observed peak flux and $B\sim 1,860 \, \rm{G}$. In addition, they obtain a maximum Lorentz factor of $\gamma_{\rm max}=3.4\times10^{6}$ by equating the SR cooling and acceleration timescales. From the spectral peak, they calculate the electron number density needed to power the emission as $n_{\rm e}=3.5\times 10^{8}\, \rm{cm^{-3}}$, noting that this far exceeds the GJ charge number density of $n_{\rm GJ}=1.1 \, \rm{cm}^{-3}$. Thus, they propose that the non-thermal emission originates far from the WD surface, rather from the region where the magnetic winds are interacting (a hypothetical high-density bow-shock region), and that the electrons are injected into the WD magnetosphere by the companion. However, the GJ charge number density only relates to the number of excess charges, not the total number of charges, meaning that the WD magnetosphere could in principle still supply the number of electrons needed to explain the SR flux.

A similar bow-shock locale for the optical and X-ray emitting particles is envisioned by \citet{Katz2017}, who speculates that the high spin-down rate of AR Sco may be indicative of it being a missing link between synchronously rotating polars and asynchronous intermediate polars. Thus, as in AE Aquirii, the dominant source of energy may be synchronisation of the orbital and WD spin periods, rather than accretion. This model attempts to explain the displacement of the observed optical maximum from the superior conjunction of the companion at $\phi_{\rm b}=0.5$. \citet{Katz2017} proposes two possible scenarios: (i) That the power dissipated by the interaction of the $B$-fields of the WD\footnote{\citet{Katz2017} estimates the WD's $B$-field strength to be $B_{\rm WD}\sim10^8$~G, comparable to that found when using a vacuum magnetic dipole field spinning down \citep{Geng2016,Buckley2017,Takata2017}.} and the companion (that forms a bow shock) is enhanced on the leading companion face compared to the trailing face, the first being relatively hotter. (ii) Alternatively, that there is a misalignment of the WD spin axis and the orbital axis, as well as the WD spin and magnetic axes, and also potential oblateness of the WD, leading to precession of the spin axis, and thus modulated dissipation, and ultimately a drift in the maximum optical flux with phase. \citet{Katz2017} lastly obtains a lower limit for the inclination angle ($\sin^{3}i \geq 0.42$) of the binary orbit using the mass function from \citet{Marsh2016}. 

\citet{Katz2017} estimated a precession period of $20 - 200$ years ($\sim2^{\circ}-20^{\circ}$ per year) for the WD, depending on the relative angle between WD spin and orbital axes and the WD oblateness. In our previous paper \citep{DuPlessis2019}, we calculated the Lense-Thirring (frame-dragging or geodetic) precession using the value of $\zeta$ that we obtained from our modelling, and assuming $\zeta\approx i$, finding a precession rate of $0.8^{\circ}$ per year for a circular orbit. Analysing archival optical data with a baseline of 100 years, \citet{Peterson2019} concluded that the precessional model proposed by \citet{Katz2017} is inconsistent with the available optical data. Conversely, a recent analysis of the X-ray light curves by \citet{Takata2020} indicated an evolution from a single-peaked structure (2016/2018 \textit{XMM}-Newton observations) to a double-peaked structure (2020 \textit{NICER} observations) when the light curves are folded at the beat frequency, one reason cited being WD precession.

\citet{Garnavich2019} find indications in their spectroscopy observations of the presence of a slingshot prominence, leading them to propose that the beat-coupled emission is due to magnetic reconnection events from the interacting magnetospheres of the WD and its companion. They find that the photon index of the optical emission changes significantly at different orbital phases and attribute this to SR cooling and the rate at which new particles are injected to replenish the old particles. \citet{Garnavich2019} further show that by implementing two existing models, the shape of the optical light curves can be reproduced. The first model suggests two heated spots on the surface of the companion, namely at the first inner Lagrangian point and at the opposite pole facing away from the WD. The second is the model by \citet{Katz2017}, discussed above. Both models can broadly capture the light curve modulation shape.

\citet{Bednarek2018} considers hybrid (lepto-hadronic) as well as pure hadronic models to account for the non-thermal X-rays detected from AR Sco, in addition to making predictions for the gamma-ray band. However, recent analysis of the 10-year-averaged \textit{Fermi} LAT data yielded only $2\sigma$ upper limits \citep{Singh2020} in the GeV band. This means that only the lowest-flux scenarios for the GeV to TeV band may still be viable.

\subsection{Magnetospheric-Origin Emission Models}
These models place the origin of the non-thermal optical and X-ray emission inside the WD magnetosphere. Details regarding the particle injection remain uncertain, but the majority of these models assume that most of the particles originate with the companion and cool via SR in the WD magnetosphere.

\cite{Takata2017} estimated a lower limit for the Lorentz factor of the electrons injected from the companion into the WD magnetosphere as $\gamma_{\rm min}=50$, based on the ratio of the magnetic dissipation power (due to the WD field sweeping across the companion surface) to the particle injection rate from the companion surface.
They next estimated the WD $B$-field at the surface of the companion as $B=195 \, \rm G$, and this yields an SR cooling time of $t_{\rm SR}=400 \,\rm{s}$, significantly exceeding the light crossing time $t_{\rm c}=a/c=2.5 \, \rm{s}$. This illustrates that the particles will produce the bulk of their SR as they move toward the WD, given that the $B $-field strength rapidly rises.
Their next step was to solve a set of coupled  transport equations similar to those of \cite{Harding2005}, used in a radio pulsar context, but they rewrote these equations in terms of the first adiabatic invariant $p_{\perp}^{2}/B$. 
Neglecting SR, they estimate the radial distance of the magnetic mirror point at which $p_{\perp}$ increases and  $p_{\parallel}$ decreases until the particle turns around, at which point $\theta_{\rm p}\sim 90^{\circ}$. In this scenario the assumptions that $\gamma\gg 1$ and the pitch angle $\theta_{\rm p}\ll 1$ are violated and a general formalism is required.
By solving the full set of equations, \citet{Takata2017} found that close to the companion, the Lorentz factor stays constant until the particle reaches the magnetic mirror, where it emits most of its energy via SR. The particle is then turned around and travels towards the opposite WD magnetic pole, until it encounters a second mirror, etc. At each mirror point, the particle emission contributes progressively less to the cumulative emission, since the particle energy rapidly declines after each turn-around. The particles are thus confined to the closed $B$-field lines until they are reabsorbed by the companion. 

\cite{Takata2019} found that only particles with small enough initial pitch angles become trapped in the WD magnetosphere.
They calculate the radiation direction and Stokes parameters to model the light curves and linear polarisation signatures of the system. They also study different initial pitch angle distributions. 

Recently \citet{Takata2020} analysed \textit{NICER},  \textit{XMM}-Newton, \textit{Chandra}, and \textit{NuSTAR} data, finding that the single-peak X-ray light curves from the 2016/2018 \textit{XMM}-newton observations seem to have evolved to a double-peaked light curve in the 2020 \textit{NICER} observation. They propose that this could be an intrinsic evolution in X-ray emission, due to changes in the emitting plasmas' temperatures, or due to the non-uniform sampling of the observations in orbital phase. Additionally, they suggest that the WD spin-coupled emission is due to particles being injected with small pitch angles and the beat-coupled emission due to particles with larger pitch angles.

\cite{Potter2018b} observed AR Sco for $\sim65$~h over two consecutive years.
Their dynamic pulse profile images (i.e., total / linear / circular flux or PPA vs.\ orbital and spin or beat phase) are extremely stable, with the same morphology occurring over one binary cycle, seen over the span of 1 year. Removal of the sideband frequencies (predominantly the beat frequency, i.e., the difference between the spin frequency $\omega$ and orbital frequency $\Omega$) yielded enhanced spin modulation at all values of $\phi_{\rm b} $; such stable structures imply emission region(s) fixed in the WD rotating frame.
\cite{Potter2018b} next applied a geometric model (i.e., invoking relative intensities for two SR components) to reproduce the polarisation observations (before removal of sideband-frequency data), namely the signatures seen in boxes a1, a2, b1, b2 of their Figure~2 (e.g., slanted lines of intensity or lack of intensity of linear flux vs.\ orbital $\phi_{\rm b}$ and WD $\phi_{\rm s}$. The symbols `a' and `b' indicate two magnetic poles, while `1' and `2' indicate primary (bright) or secondary (dim) emission. They were unable to qualitatively reproduce these structures when invoking emission sites fixed to the binary frame (e.g., the companion face or bow shock). Conversely, a scenario where emission regions are locked to the WD frame seems plausible.
They assumed that relativistic particles are injected from the companion as the open magnetic field lines of the WD sweep across the surface of the companion. These particles travel along the magnetic field lines towards to the nearest WD magnetic pole, while some particles are thought to be travelling to the opposite pole. As the particles approach the pole, SR cooling becomes more dominant, given the rise in magnetic field strength, causing the particles to radiate most of their energy near putative magnetic mirror points. As an example scenario, they set the WD magnetic inclination angle, the system inclination angle, and a beaming (half-opening) angle for the SR to $40^{\circ}$, $60^{\circ}$, and $45^{\circ}$, respectively.  
The emission is beamed in the direction of the particle's motion (SR yielding a double-peaked profile for linearly and single-peaked profile for circularly polarized emission; corresponding to `split' and single observed emission components).
Based on the peak flux intensities observed at the beat-phase range of $0.9 - 1.1$, they invoked enhanced emission at $\phi_{\rm b}=0.7 - 0.9$. Given the fact that the particles are in-flowing, this will lead to an enhancement for an observer at $\phi_{\rm b}= 0.2-0.45$ (where $\phi_{\rm b}=0$ corresponds to inferior conjunction of the companion). 
This is attributed to enhanced particle injection at those phases where field lines from the WD magnetic poles sweep across the companion surface.

In our previous work \citep{DuPlessis2019}, we implemented a standard geometric RVM, typically used to fit radio PPA data from canonical pulsars, and applied the model to optical PPA data from the AR Sco WD \cite{Buckley2017}. These PPAs resulted from binary phase-averaged emission (including normalized binary phases $\phi=0.07 - 0.23$). This allowed us to constrain $\alpha$ and $\zeta$ of the WD, as well as its mass. Being a geometric model, it makes no pronouncement on the origin of the injected particles, but rather utilizes that $B$-field structure of the WD to explain the PPA sweeps seen in optical range. Due the large duty cycle of the optical PPA, we were able to find well-constrained parameter values compared to standard radio pulsar fits, namely $\alpha ={86.5^{\circ}}^{+3.1}_{-2.8}$ and $\zeta ={60.4^{\circ}}^{+5.4}_{-6.0}$. We were also able to constrain the mass of the WD mass as $m_{\rm WD} =1.00^{+0.16}_{-0.10} \, M_{\odot}$ when assuming that the binary inclination $i=\zeta$, yielding mass constraints that fall within the constraints derived by \cite{Marsh2016}. Since the RVM does not specify if the particles are moving towards or away from the polar cap, our model is reconcilable with the postulate that particles are injected from the companion into the magnetosphere of the WD as suggested by \citet{Takata2017,Potter2018b,Lyutikov2020}.

\citet{Singh2020} propose two locales containing relativistic electrons that radiate SR: one near the interacting stellar winds and one closer to the WD surface. By fitting the broad-band SED using broken-power-law injection spectra, they infer radii for the purported emission regions, as well as the local magnetic field strength at each location. \citet{Singh2020} use both these SR components to fit the observed X-ray spectrum. Comparing the SED from \cite{Takata2018} and some of our own SED fitting, it seems that \citet{Singh2020} fit their SR spectrum to the thermal X-ray component instead of the non-thermal component. It is furthermore not clear what the model degeneracies are, since this may not be a unique solution.

\subsection{Intermediate Polar Model}
Similar to the previous class of models, this model also suggests emission originating from the WD magnetosphere (and particles coming from the companion) but relies on AR Sco being an intermediate polar.

\citet{Lyutikov2020} depart from a strict isolated-pulsar-like spin-down scenario where the magnetic field is dependent on the WD's spin period and time derivative thereof, since there seems to be a lack of coherent radio emission and AR Sco is not an isolated WD (and no isolated WD produces pulsed coherent radio emission). Furthermore, the usual electron-positron pair production envisioned in pulsars is associated with $\gamma$-ray emission, but no such emission has been detected from AR Sco. 
The scenario \citet{Lyutikov2020} favor is rather that AR Sco is a most peculiar cataclysmic variable - a transient propeller system similar to AE Aquarii (a highly asynchronous intermediate polar), despite the lack of spectroscopic evidence for an accretion column and presence of polarised emission in this system that point to the contrary. 

The model by \citet{Lyutikov2020} was proposed to motivate the current short spin period of the WD in AR Sco, which might indicate a past episode of fast accretion. They calculate the mass-loss rate from the companion needed to spin up the WD to its current spin period.
Since the mass-loss rate has a dependence $\dot{M}\propto B_{\rm WD}^{2}$, they find that for large magnetic fields, the required mass-loss rate is unreasonably high. Their model thus predicts that the surface magnetic field strength of the WD is $B\sim5\times 10^{5}-5\times10^6\,\rm G$, assuming accretion onto the WD. Since this magnetic field is low compared to that inferred by \cite{Buckley2017}, they obtain a more reasonable mass loss rate needed to spin up the WD to its current spin period. However, accretion is not the only way of spinning up a WD. For example, WDs are born with slow periods, and having a long lever arm, this facilitates large angular momentum transfer rates via clumpy mass flow \citep{King1993,Wynn1995}. Alternatively, mass ejection could have taken place during a nova explosion, spinning up the WD. Also,  coupling between the magnetic fields of the two binary members may facilitate spin-up. These alternative scenarios may allow for larger WD magnetic fields.

Within the plasma-loading scenario of \citet{Lyutikov2020}, the red dwarf ejects relatively slow-moving, semi-ionised plasma along its magnetic field lines directed towards the WD. The plasma penetrates the WD magnetosphere and may be further ionised by UV photons from the WD surface (or other non-thermal photons), leading to non-thermal emission and magnetic bottling (mirroring), similar to what was proposed by \cite{Takata2017,Potter2018}. Using WD atmospheric models,  
\citet{Lyutikov2020} extract an upper limit of $T\sim12,000\, \rm {K}$ for the effective WD surface temperature; this is comparable to $T=9,750\, \rm {K}$ found by \cite{Marsh2016}. Thus, UV radiation from the WD dwarf surface and non-thermal emission in the WD magnetosphere may ionise the purported inflowing plasma, potentially causing such plasma to be flung away by the rotating magnetic fields as the system enters a propeller state. As the plasma stops penetrating the WD magnetosphere, the flux of the non-thermal components drops and the plasma becomes less ionised and starts penetrating the WD magnetosphere once again, thus resulting in a self-regulating process of quasi-periodic emission. 
However, while an accretion disc requires a low magnetic field, it is inconceivable that such a low  field will be effective enough to establish a propeller effect; a disc would develop in the absence of the former, but then this would lead to strong observable signatures such as double-peaked emission lines, given the orbital inclination, which are not detected (no broad emission lines have been; rather, those seen are connected to the face of the secondary). Thus the assumption of a low magnetic field with mass flow from secondary, and propeller scenario due to ionization of the mass flow is questionable.

In their first-order calculations (neglecting particle dynamics and pitch angle evolution), \citet{Lyutikov2020} propose that magnetic reconnection takes place at the location of interacting magnetic fields. Particles are then accelerated and radiate in a local magnetic field of $\sim 2\times 10^{3} \, \rm G$, where the resulting potential difference is enough to accelerate particles to relativistic energies. Their model assumes that the optical and X-ray non-thermal components originate from the WD magnetosphere, taking a similar approach to that of \cite{Takata2017}. However, they do not include a prediction for the SED, light curve, or polarisation. The reconnection events are believed to be variable, given the changing orientation of the companion's magnetic moment relative to the WD, and this may be responsible for the modulation in the observed non-thermal flux at the WD’s rotation and beat periods. Radio emission is thought to be produced in the companion magnetosphere, and \citet{Lyutikov2020} do not expect detectable emission in the TeV band, given the relatively low soft-photon energy density compared to the magnetic energy density. However, it is unclear how particle acceleration via magnetic reconnection may operate to yield the observed multi-wavelength spectrum. If the plasma is denser close to the WD, this may inhibit formation of extended current sheets where efficient electron acceleration may take place.

More recently, \citet{Garnavich2020} reanalyzed HST data by calculating the average spectrum of the minima between SR pulses. When averaged around inferior conjunction, they found evidence for a quasi-molecular hydrogen absorption band, inferring a WD surface temperature of $11,500 \pm 500 \rm{K}$. Additionally, they infer the presence of two hotspots near the magnetic poles of the WD, ranging between $23,000~\rm K$ and $28,000~\rm K$.
They argue that while these hotspots may be explained energetically by SR irradiation of the WD surface \citep{Potter2018}, the dominant emission from the region of the magnetic mirrors may miss the WD surface. Thus, they discard this possibility to explain the existence of the hotspots. However, one has to consider that other combinations of initial pitch angles and lepton Lorentz factors (prior to the mirror, or penetration of the mirror by small-pitch-angle particles) may yield irradiation of almost the entire WD surface. 
Building on the model by \citet{Lyutikov2020}, \citet{Garnavich2020} propose that protons trapped in the WD magnetosphere may escape and heat the WD polar caps. 
However, evidence for 15,000 K gas on the irradiated face of the secondary, though, implies further investigation of the interaction mechanism between the WD and the secondary. Furthermore, the high temperatures of the hotspots near the WD poles would imply ionization of the gas originating from the secondary before it reaches the WD's inner magnetosphere, implying some further revision to the original picture of \citet{Lyutikov2020}. \citet{Garnavich2020} lastly note that the absence of strong Zeeman splitting implied an upper limit of 100~MG for the surface magnetic field of the WD. 

\section{Conclusion}\label{sec:6}
\begin{figure}
\centering
\includegraphics[width=20pc]{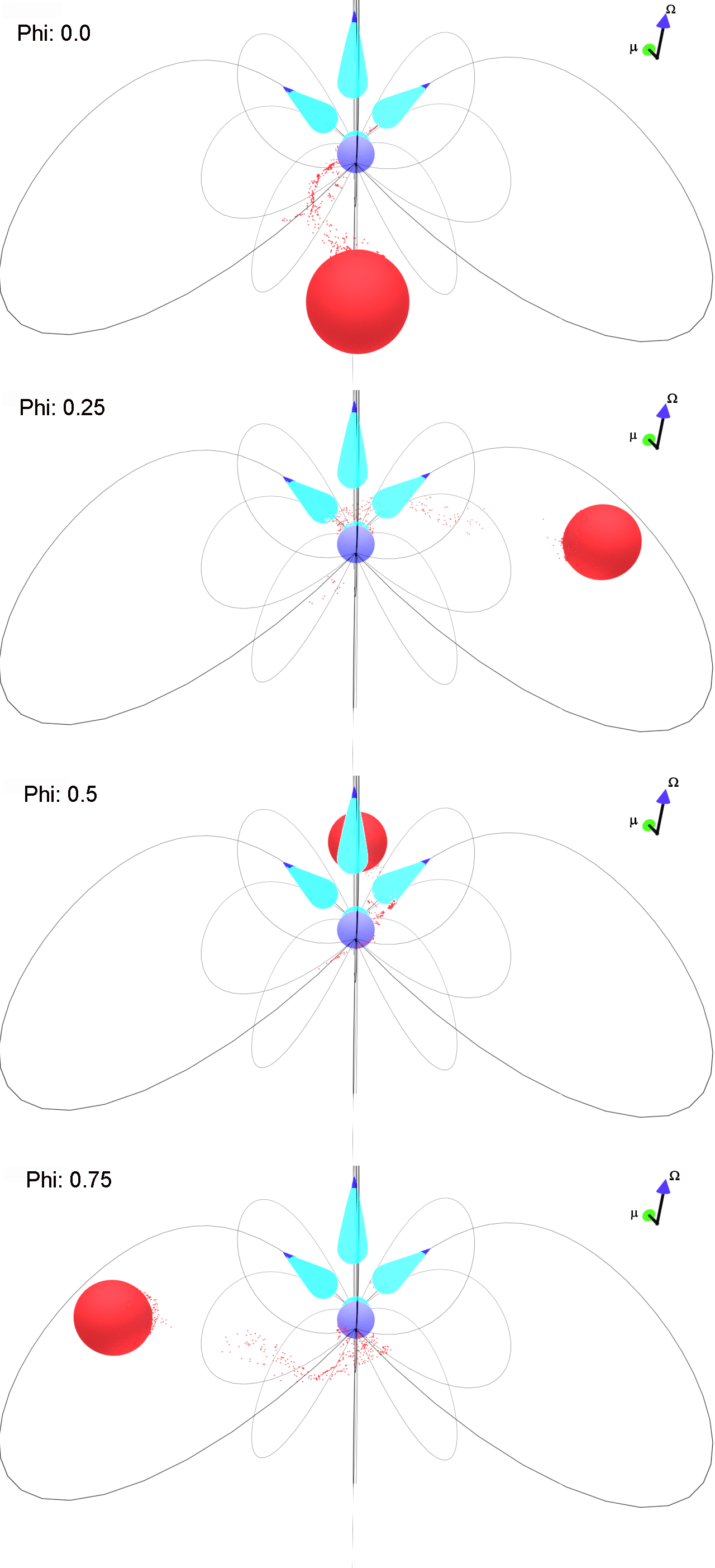}
\caption{An illustration of AR Sco at different $\phi_{\rm b}$. The blue arrow shows the direction of the spin vector and the green arrow the direction of the magnetic moment of the WD. The cyan cones show the emission of the electrons pointing towards the observer as the infalling electrons move towards the WD's polar cap. The observer's line of sight is aligned to how the illustration is viewed (perpendicular to the page).}
\label{Sys_Illus}
\end{figure}

In this paper, we fit a standard RVM to a much more extensive set of phase-resolved optical polarization data on AR Sco. We found adequate fits to the PPA curve over most of the orbital period. Our new fits to the data pulsed at the WD spin period point to a WD magnetospheric origin of the emission. 
We find a $\sim 10^{\circ}$ variation in $\alpha$ and $\sim 30^{\circ}$ variation in $\zeta$ over the orbital period. The variation patterns in these two parameters are robust, regardless of the binning and epoch of data used. Below, we consider various scenarios for this variation. 
Figure~\ref{Sys_Illus} is an illustration of the system, defining $\phi_{\rm b}=0$ as inferior conjunction of the companion. Emission cones of the incoming particles as they point towards the observer are shown in cyan color.

\begin{figure}
\centering
\includegraphics[width=20pc]{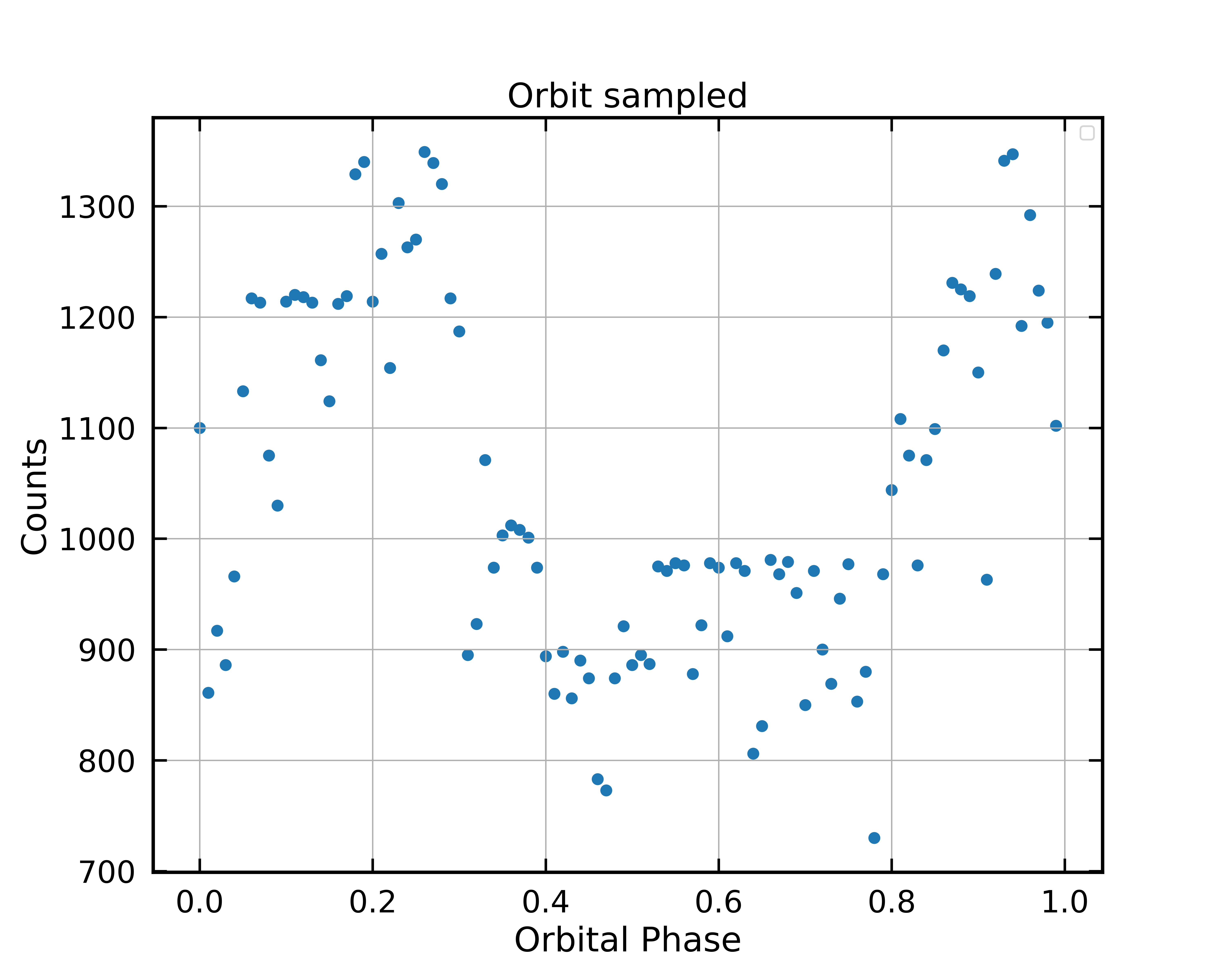}
\caption{Observational data point counts vs.\ orbital phase for the observations of \citet{Buckley2017}.}
\label{Orb_sample}
\end{figure}

First, this variation could be due to precession of the WD as first proposed by \citet{Katz2017} to explain the optical light curve shape, and also by \citet{Takata2020} based on the temporal evolution of the X-ray light curves. Second, this variation may be due to the inhomogenous sampling of data over the orbital period. For example, if particle injection from the secondary occurs at particular phases \citep{Potter2018}, the light curve shape could be biased based by observation cadence. Figure~\ref{Orb_sample} shows that for the data we considered, the orbit was relatively uniformly sampled, except around orbital phase $0$ and from $0.3-0.8$. 
Third, the emission source could be wobbling or may be asymmetrically sampled by our line of sight. Fourth, different particle populations may be injected with different original pitch angle distributions \citep{Takata2020}. In their $Z^{2}_{2}$-periodogram analysis, \citet{Takata2020} found that at orbital phase $0.1-0.6$, the spin frequency and its harmonics are as strong and stronger than the beat frequency and its harmonics, but at orbital phase $0.6-1.1$ there is almost no presence of spin frequency and its harmonics. We find a similar trend in the first panel of Figure~\ref{FFT_orbital_2bins}  for orbital phase $0.1-0.6$, but in the second panel for $0.6 - 1.1$. Contrary to \citet{Takata2020}, our spin component is still present but at a low level\footnote{The reason the spin frequency is still present in the second phase region in the optical compared to the X-rays could be due to the fact that the optical data are of higher quality and the non-thermal component is much stronger in the optical}. A Lomb-Scargle analysis for orbital phase $0.3-0.5$\footnote{This is the region \cite{Potter2018} suggest particles are injected from the companion and where the maximum linear flux occurs.} as shown in the last row of Figure~\ref{FFT_orbital_2bins}, reveals that the power of the first harmonic of the spin frequency is similar to that of the beat frequency, but for the second harmonic the beat frequency power is still a bit higher. These observations clearly show a mixture of spin-coupled and beat-coupled emission, with the spin period becoming more dominant at certain orbital phases. Similar to \cite{Takata2020}, we thus suggest that there could be two populations of particles with different initial pitch angle distributions or different injection rates at slightly different phases, and both populations cool via SR. \citet{Takata2020} propose that particles with smaller initial pitch angles generate the spin-coupled emission since they penetrate deeper into the WD magnetosphere and have more beamed emission cones. The particles with larger initial pitch angles are proposed to generate the beat-coupled emission, since the emission may be observed over a larger region of the orbit. This could lead to the variation of $\zeta$ found in Figure~\ref{Orb_fit}, as different emission regions are sampled at different orbital phases. Fifth, a mix of WD and companion emission might have an effect on orbitally-dependent depolarization. Lastly, changes in the magnetic field structure due to the interaction with the companion may account for some variability in the signal, but it is unclear if such structural changes can happen on the time scale of one orbital period.

An outstanding question within the framework of a model that suggests particles are injected from the companion is what happens to the particles once they have cooled. To address this problem fully, a more sophisticated model (beyond the RVM) is required, thus we will only speculate here. First, a fraction of injected electrons (and possibly protons) that are not mirrored, may be absorbed by the WD (whose surface may thus be heated). Additionally, the companion may also reabsorb mirrored charges. Second, some of the particles may be flung into the WD's wind and escape from the WD's magnetosphere. Third, if both electrons and protons are injected into the WD magnetosphere, the protons will not cool as efficiently. This could lead to charge separation or build up of charge, implying an increased optical depth for absorption of SR. Such a charge build-up may also impact the injection of new particles as well as the current flow patterns as well as the maximum voltage experienced by charged particles. Furthermore, an electric field will be induced that will help to ameliorate such a build-up.

To test these ideas, we intend to implement a more sophisticated emission model in future, solving the particle dynamics to generate phase-resolved light curves and SEDs. Continued multi-wavelength observations may further contribute to constrain the properties of this intriguing source.

\section*{Acknowledgements}
This work is based on the research supported wholly / in part by the National Research Foundation of South Africa (NRF; Grant Number 99072). The Grantholder acknowledges that opinions, findings and conclusions or recommendations expressed in any publication generated by the NRF supported research is that of the author(s), and that the NRF accepts no liability whatsoever in this regard. Z.W. is thankful for support from the NASA Postdoctoral Program. This work has made use of the NASA Astrophysics Data System.

\section*{Data Availability}
The data used in this article is the polarimetric data from \citet{Potter2018b}. The authors of the article should be contacted to obtain the archived data. Any of the data analysis, RVM, and MCMC-fitting code used for this article will be shared upon reasonable request.

\appendix
\section{Extra Plots}
In this Appendix, we show the Lomb-Scargle power spectra for different segments of orbital phase, as summarized in Figure~\ref{Lomb_amp}. The left column of each panel in both figures show the first harmonic of the spin and beat frequencies, the middle column shows the second harmonics, and the third column the third harmonics. In these plots some of the side-band frequencies where included.
\begin{figure}
\centering
\includegraphics[width=\columnwidth,trim=0 0 0 1cm]{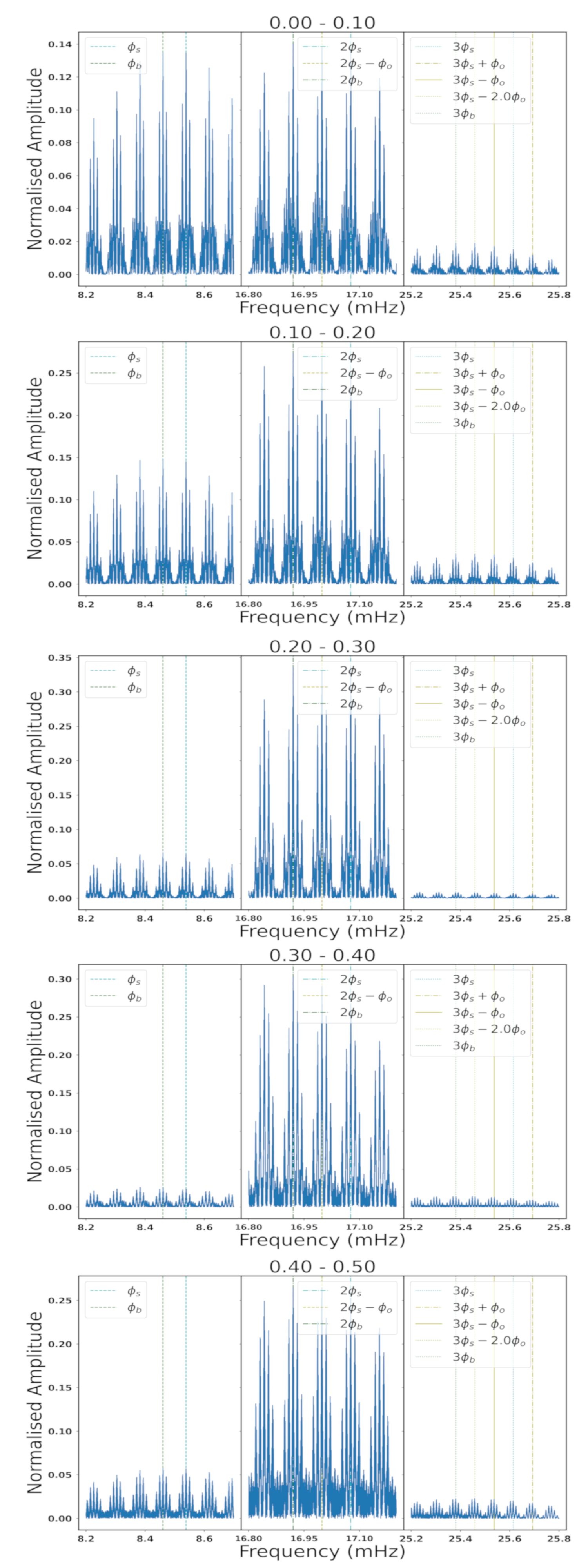}
\caption{Lomb-Scargle power spectrum for different segments in orbital phase namely $0.0-0.5$. In each panel the beat frequency and its harmonics are indicated in green, the spin and its harmonics in cyan, and other notable side-band frequencies in yellow.}
\label{FFT_orbital1}
\end{figure}
\begin{figure}
\centering
\includegraphics[width=\columnwidth,trim=0 0 0 1cm]{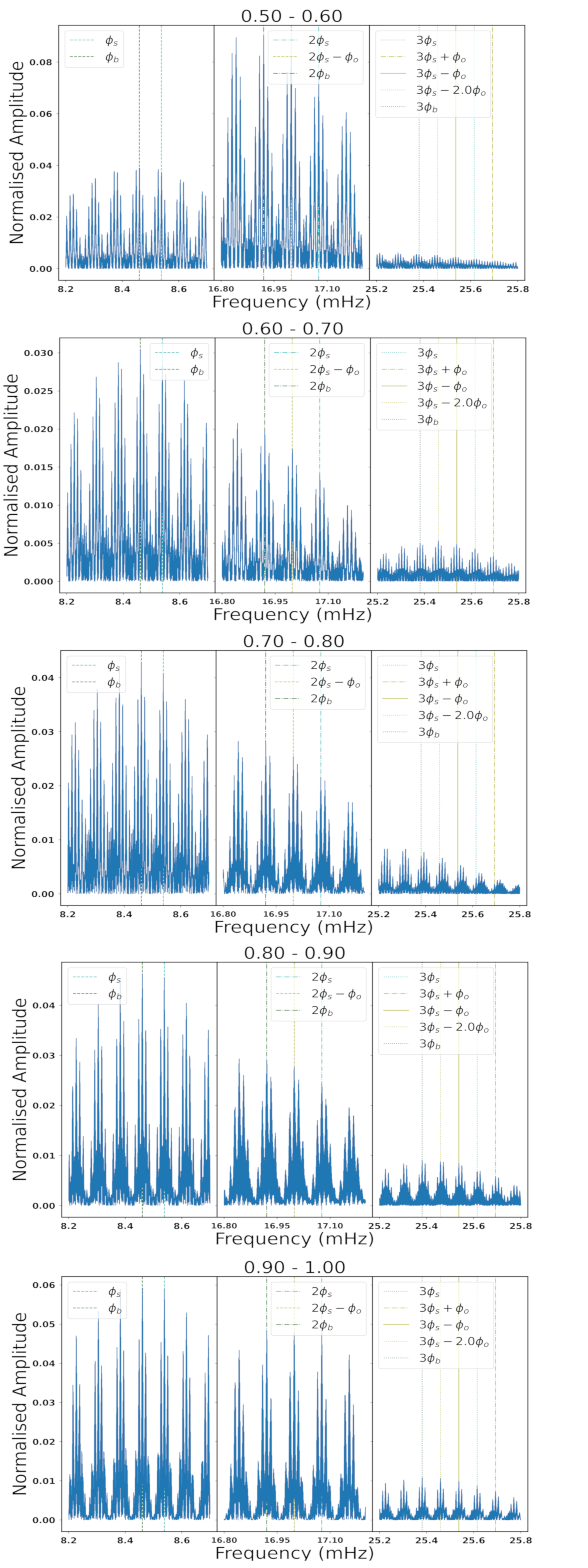}
\caption{The same as Figure \ref{FFT_orbital1} but for orbital phases $0.5-1.0$}
\label{FFT_orbital2}
\end{figure}

\bibliographystyle{mnras}
\bibliography{ARsco_refs}
\bsp	
\label{lastpage}
\end{document}